\documentclass[journal]{IEEEtran}

\newcommand{\norm}[1]{\left\lVert#1\right\rVert}

\usepackage{graphicx,epsfig}
\usepackage{caption}
\usepackage{color}
\usepackage{subcaption}
\usepackage{cite}
\usepackage{stfloats}
\usepackage{float}
\floatstyle{plaintop}
\restylefloat{table}
\usepackage{amsmath}
\usepackage{amsfonts,helvet}
\usepackage{mathrsfs}
\usepackage{algorithm}
\usepackage{algorithmic}
\usepackage{amsthm}
\usepackage{multirow}
\makeatletter
\adddialect\l@ENGLISH\l@english
\makeatother

\makeatletter

\let\proof\@undefined
\let\endproof\@undefined
\makeatother

\newcommand{\xyplane}{$x\,y$\nobreakdash-plane}
\newtheorem{theorem}{Theorem}

\newtheorem{algo}[theorem]{Algorithm}

\newtheorem{proposition}[theorem]{Proposition}

\newcounter{proposition}
\begin{document}
\title{RF Lens-Embedded Massive MIMO Systems: Fabrication Issues and Codebook Design}

\author{Taehoon~Kwon,~\IEEEmembership{Student Member,~IEEE},
        Yeon-Geun~Lim,~\IEEEmembership{Student Member,~IEEE},
        Byung-Wook~Min,~\IEEEmembership{Member,~IEEE}, and
        Chan-Byoung~Chae,~\IEEEmembership{Senior Member,~IEEE}\\

\thanks{T. Kwon, Y.-G. Lim, and C.-B. Chae are with the School of Integrated Technology, Yonsei University, Korea (E-mail: \{th\_kwon, yglim, cbchae\}@yonsei.ac.kr). B.-W. Min is with the Department of Electrical and Electronic Engineering, Yonsei University, Korea (E-mail: bmin@yonsei.ac.kr).}
\thanks{This work was in part supported by the MSIP (Ministry of Science, ICT and Future Planning), Korea, under the ``IT Consilience Creative Program" (IITP-2015-R0346-15-1008) supervised by the IITP (Institute for Information \& Communications Technology Promotion) and ICT R\&D program of MSIP/IITP [B0126-15-1017]. Part of this work was presented in~\cite{kwon2015rflens}.
 }}

\maketitle 


\begin{abstract}
In this paper, we investigate a radio frequency (RF) lens-embedded massive multiple-input multiple-output (MIMO) system and evaluate the system performance of limited feedback by utilizing a technique for generating a suitable codebook for the system. We fabricate an RF lens that operates on a 77~GHz (mmWave) band. Experimental results show a proper value of amplitude gain and an appropriate focusing property. In addition, using a simple numerical technique--beam propagation method (BPM)--we estimate the power profile of the RF lens and verify its accordance with experimental results. We also design a codebook--multi-variance codebook quantization (MVCQ)--for limited feedback by considering the characteristics of the RF lens antenna for massive MIMO systems. Numerical results confirm that the proposed system shows significant performance enhancement over a conventional massive MIMO system without an RF lens.

\end{abstract}


\section{Introduction}

\IEEEPARstart{F}{or} decades, researchers have studied wireless multiple-input multiple-output (MIMO) systems in an effort to provide higher capacity gains and better link reliability \cite{goldsmith2003capacity}. Attempts to improve MIMO techniques were first conducted with single-user (SU) setups, later evolving into multi-user MIMO (MU-MIMO) systems. A transmitter in a MIMO system has to acquire channel state information (CSI) to provide beamforming gains in SU-MIMO systems, and multiplexing gains in MU-MIMO systems \cite{gesbert2007shifting}. To achieve the theoretical bound of the MIMO broadcast channel where multiple antennas are deployed at both transmitter and various receivers, \cite{spencer2004zf,cbchae2008cbf} proposed simple zero-forcing (ZF)-based linear algorithms with limited feedback. The authors in \cite{chae2009optimality} proved that a simple linear beamforming technique, referred to as coordinated beamforming, could asymptotically reach the sum capacity performance of dirty paper coding (DPC). 

To support the exponentially increased mobile data traffic of the present, a need has now arisen for new techniques that go beyond conventional MIMO systems; such techniques must be able to provide at least 1,000 times more capacity gains than current systems~\cite{andrew20145g}. 
Among the various candidates for such a technique, a massive MIMO has been considered a promising one for 5G wireless communication systems~\cite{marzetta2010noncooperative, rusek2013scaling, ngo2011energy, boccardi20145g}. By using a large amount of antennas (64 or more) at the base stations (BSs), a significant improvement in the network capacity and energy efficiency using ZF or maximum ratio transmission/combining (MRT/MRC) can be expected~\cite{huh2012notsolarge,yang2013conzf,smallcell2015,ygcell}. With inaccurate CSI, however, the sum rate performance of a massive MIMO system may be saturated \cite{jindal2006mimo}. Therefore, an efficient channel-training and feedback method has to be carefully designed in massive MIMO systems. While the benefits of lower training overhead are more evident in a time division duplexing (TDD) massive MIMO system~\cite{jose2011tdd}, most commercial cellular systems use the frequency division duplexing (FDD) mode, which offers more benefits than the TDD mode, especially in macro-cell environments~\cite{choi2014memory}. In FDD massive MIMO systems, channel reciprocity does not hold and the receiver has to feed-back CSI to construct precoding vectors \cite{jiang2015fdd}. To maintain a certain level of CSI quantization loss, increasing feedback overhead is inevitable \cite{love2008feedback}. Consequently, massive MIMO systems, with their very large antenna arrays at the BSs, struggle to provide feedback without compression. Researchers have tried to alleviate this feedback overhead problem. They have tested channel quantization methods, including random vector quantization (RVQ) \cite{jindal2006mimo}, Grassmannian codebook \cite{love2003grassmannian}, and its extended version in time-varying channels \cite{takao2009}. These can support a large number of users. The authors in \cite{nam2012} proposed a feedback reduction technique that uses spatial correlation of users, while the authors in \cite{choi2013trellis} proposed noncoherent trellis-coded quantization for FDD massive MIMO systems. In addition, a low complexity compressive sensing-based method was also proposed in correlated massive MIMO channels~\cite{lim2014compressed, sim2015compressed}.


 What amounts to a novel approach is using, in massive MIMO systems, the radio frequency (RF) lens. Lenses are basically phase-shifting devices. They convert a divergent wavefront from a point source into a plane wave and vice versa. 
 By providing high gain, narrow beamwidth and low sidelobes in different directions, the RF lens has been fully utilized in applications such as radars and satellite communication systems~\cite{antennabook}. Traditional design of lens antenna has consisted of an array of antennas used with variable-length transmission lines to create the aperture phase profile \cite{hollung1997, popovic1998, lau2010}. Another promising antenna design that uses RF lenses with an antenna array in a MIMO system was considered in \cite{sayeed2011capmimo, sayeed2013beamspacemimo, zhang2014lens, zhang2015, kwon2015rflens}. Proposed in \cite{sayeed2011capmimo} is a discrete lens array (DLA) structure-based continuous aperture phased (CAP) MIMO at the mmWave (millimeter-wave, 60--100~GHz, region). DLA behaves as a convex lens, transferring the signals towards different points of the focal surface. Moreover, DLA narrows the beamwidth, which is still preserved in the reduced RF chain operation, making it possible to reduce the power consumption and interference between the streams. In their subsequent  work~\cite{sayeed2013beamspacemimo}, the authors demonstrated that by combining the concept of beamspace MIMO communication with a hybrid analog-digital transceiver, CAP-MIMO achieves near-optimal performance with dramatically lower complexity. The studies in \cite{sayeed2011capmimo, sayeed2013beamspacemimo} mainly concern line-of-sight (LOS) mmWave channels, where only a short transmission range is possible. The authors in \cite{zhang2014lens} suggested alternative antenna designs that use an electromagnetic (EM) lens in front of a large antenna array at the BS. In the single-cell multi-user uplink setup of massive MIMO cellular systems, they demonstrated performance gains and a reduction of signal-processing complexity. They considered the uplink setup with channel estimation through uplink training. In our prior work \cite{kwon2015rflens}, we considered an RF lens-embedded massive MIMO system at a single-cell multi-user downlink setup. We introduced the BPM to estimate the power profile of the signal. Whether the measured results agreed with our BPM-based simulations, however, was not considered by the system. In addition, lens fabrication issues have not been addressed in prior work.

 
In this paper, extending the results from \cite{kwon2015rflens}, we study an RF lens-embedded massive MIMO system at the single-cell multi-user downlink setup, proposing a codebook design capable of enhancing the performance gain. First, we analyze the RF lens-embedded massive MIMO systems and create a new channel model. Since the channel model is directly related to the signal power profile in the proposed system, theoretical values should be accurately estimated. We adopt the BPM to calculate the propagating and focusing of incident beams controlled by the RF lens. We also investigate more details about the calculating phase profile in front of and behind the lens surface. Since the BPM simulation in \cite{kwon2015rflens} was based on a thin lens approximation, a more realistic analysis had to be conducted, one that would include a real experimental setup.
 We therefore provide measurement results and demonstrate their accordance with our analysis. Using a power profile vector acquired by BPM in designing the new codebook for the $k$-th user enables the quantized channel to be associated with the $k$-th user's power profile vector. This way, the channel variance of each antenna component satisfies the multi-variance nature of our new channel model. This proposed codebook in the RF lens-embedded massive MIMO systems is termed  multi-variance codebook quantization (MVCQ). We also propose several sub-algorithms that can reduce the computational complexity without notably degrading performance.

\smallskip
The main contributions of this paper are as follows:
\begin{itemize} 

\item We investigate an RF lens-embedded massive MIMO system for FDD operation. We design a new codebook--the multi-variance codebook quantization (MVCQ)-- for limited feedback by considering the characteristics of the RF lens antenna. We evaluate the proposed system in realistic environments using a 3D-ray tracing tool-based system level simulator. To the best of our knowledge, the FDD lens-embedded massive MIMO operation has yet to be studied, permitting the continuation of the feedback overhead problem of massive MIMO systems. 

\item We provide numerical analysis based on Fourier optics and BPM to calculate the additional power factor captured by each antenna. The field distribution of the RF lens is calculated through lens geometry analyses under several assumptions. We adapt BPM in the proposed codebook.

\item We fabricate an RF lens that operates on a 77~GHz (mmWave) band and discuss fabrication issues. By fabricating a parabolic dielectric lens with various materials, we measure the radiation patterns and the focusing intensity. We provide simulation results of BPM and demonstrate its accordance with measurement results. Most prior work on lens antenna arrays has paid little attention to fabrication and design issues, and researchers have yet to develop a method that offers insight into constructing a system model of lens-embedded systems. 

\end{itemize}

  We anticipate our contributions to yield a wide range of insights for RF lens-embedded massive MIMO systems regarding the practical fabrication issues and performance analysis.
 
This paper is organized as follows.
Section II contains a brief overview of our proposed RF lens-embedded massive MIMO system. Basic configurations and system parameters are defined. Section II also details the process of constructing the channel model.
Section III describes the calculating process of the phase function by using the lens geometry. The section also introduces algorithmic details of the BPM. 
In Section~IV, we propose a channel feedback algorithm based on our lens-embedded channel model. The BPM and other methods are considered to calculate the additional power coefficients in our new codebook construction model.
Section V demonstrates the fabrication strategy based on theoretical results, and presents the final antenna design and the measurement results.
Numerical results in 2D and 3D environments are shown in Section VI. Finally, Section VII presents our conclusions and proposes future work.\footnote{Throughout this paper, we use upper and lower case boldface to describe matrix $\pmb{A}$ and vector $\pmb{a}$, respectively. The transpose and the Hermitian transpose of a matrix is notated as $(\cdot)^T$ and $(\cdot)^*$, respectively. Other notations are explained where they are used.}

\section {System Model}
\label{model,background}

\subsection{System Overview}
\label{Overview}	
\begin{figure}[!t]
	\setlength{\belowcaptionskip}{-15pt}
  \centerline{\resizebox{1.0\columnwidth}{!}{\includegraphics{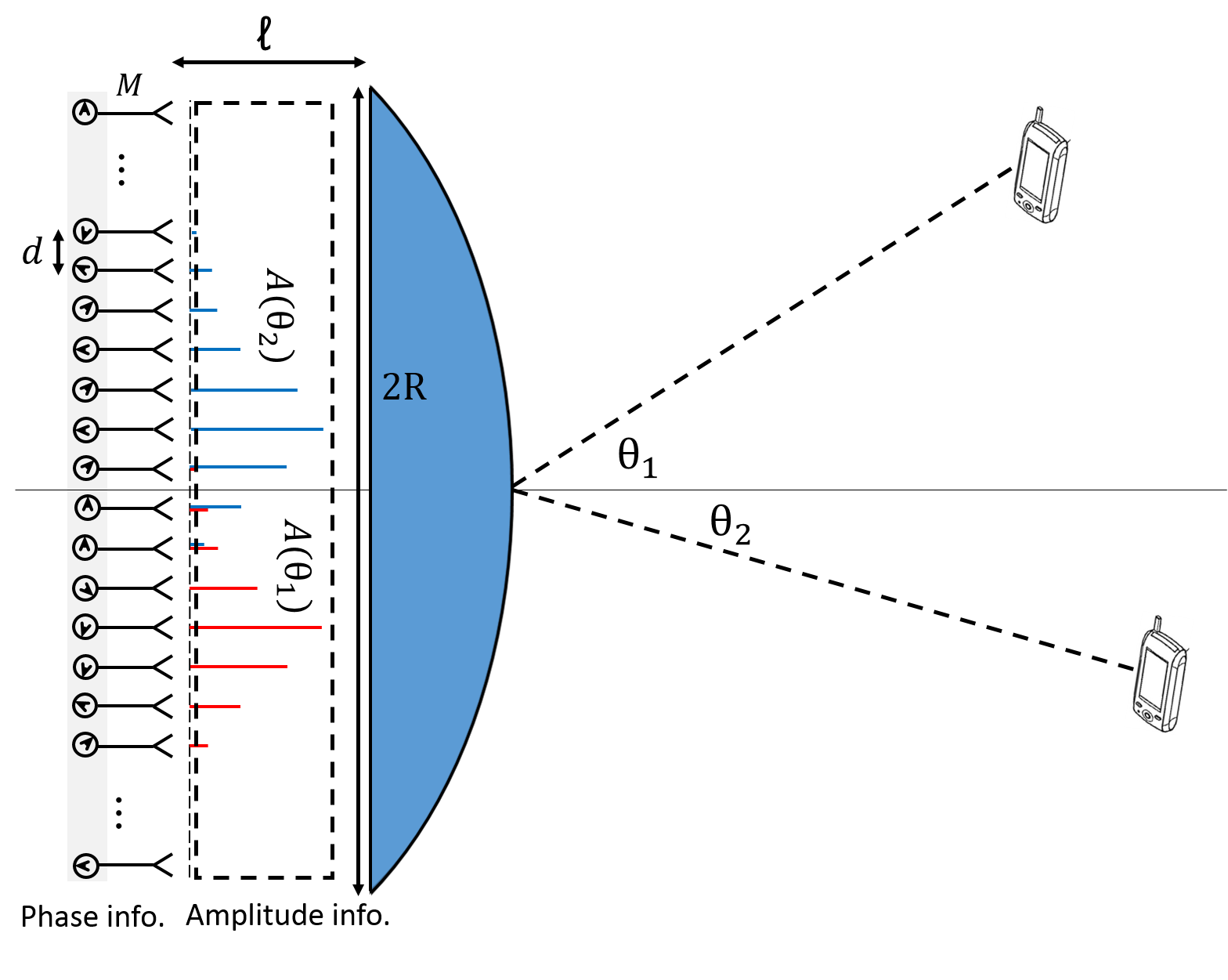}}}
   \caption{A schematic of the proposed RF lens-embedded massive MIMO system.}
   \label{Fig:1D_lensmodel}
\end{figure}

We consider a single-cell multi-user downlink system where a large antenna array is equipped with the RF lens at the BS. Figure~\ref{Fig:1D_lensmodel} offers a schematic of the proposed RF lens-embedded massive MIMO system.
The BS antenna can be divided into two parts--a large antenna array part and an RF lens part. For the antenna array, we first need to determine $d$ (the separation of adjacent antennas) and $M$ (the number of antennas at the BS). System parameters $f$ (the focal length of the lens), $D$ (the aperture diameter of the RF lens) and $\epsilon_r$ (the electric permittivity of lens material) should also be determined for the RF lens part. $f$, $D$, $\epsilon_r$, and contours of classical lenses can be described with a certain formula. To be specific, if the values of focal length $f$, and the diameter of the lens, $D$, are fixed, other contour parameters must be determined to fulfill the focusing property of the RF lens. More details are included in Section V-A. The parameter $\ell$ (the distance between the lens and the antenna arrays) is also important, as it can determine the focusing level of the RF lens.\footnote{Note that there might be a gap between the geometrical focal region and the real focal region of the system; thus the parameter $\ell$ is empirically determined in this paper.}  For array configurations, 1-dimensional (1D) uniform linear array (ULA), 2D uniform square array (USA) and circular array or even 3D antenna array configurations can be evaluated. In this paper, for simplicity, we only consider the 1D ULA.\footnote{Future work will consider higher dimensional antenna array configurations.}

 
\subsection{Channel Model}
\label{Model}
We first consider the channel without an RF lens. As noted above, we will restrict our attention to MIMO channels with an ULA of $M$ antennas at the BS and one antenna at each user. 
Here, we adopt the spatially correlated MIMO channel model. Two popular approaches include parametric models (PMs) and nonparametric models (NPMs), which are widely used to simulate the correlated MIMO channels \cite{ertel1998corr}. By choosing a suitable correlation matrix at the transmitter side, we adopt the Kronecker model, which is NPMs, to describe each channel vector between the BS and the $k$-th user~\cite{forenza2007corr}. 
 Let $\theta_{k}$ be the angle where the $k$-th user is located. We assume a Laplacian distribution for the angle of departure (AoD), the same as the power angular spectrum (PAS) in a 2D spatial channel model~\cite{release12}. The PAS can be expressed as the truncated Laplacian probability density function (PDF) as
  \begin{displaymath}
   P_{\theta}(\theta) = \left\{
     \begin{array}{lr}
       \frac{\beta}{\sqrt{2}\sigma_{\theta}}e^{-|\frac{\sqrt{2}\theta}{\sigma_{\theta}}|} & : \theta \in [-\pi,\pi)\\
       0 & : \text{otherwise}
     \end{array}
   \right.
\end{displaymath} 
where $\theta$ is a random variable describing the offset of the mean angle for the $k$-th user, $\theta_{k}$, $\sigma_{\theta}$ is the standard deviation of the PAS, and $\beta$ is the normalization factor. 
In the NPMs, the coefficients of a spatial correlation matrix at the BS, given as an $M \times M$ matrix $\pmb{R}_{k, {\mathrm{TX}}}$, are described by a certain angle spread and AoD of the $k$-th user. Closed-form expression for the correlation coefficients between the $i$-th and the $j$-th antennas is derived for the Laplacian azimuth angle distribution~\cite{forenza2007corr}.
\begin{eqnarray}
\left[\pmb{R}_{k, {\mathrm{TX}}}(\theta_{k},\sigma_{\theta})\right]_{i,j} = \frac{\beta e^{j\kappa d(i-j)\sin{\theta_{k}}}}{1+\frac{{\sigma_{\theta}}^2}{2}[\kappa d(i-j)\cos{\theta_{k}}]^2}.\nonumber \\ 
(k = 1,2,\cdots,K) \nonumber 
\end{eqnarray}
Therefore, the spatially-correlated MIMO channel between the BS and the ${k}$-th user, which is an $M \times 1$ column vector, is given by
 \begin{equation}
 \pmb{h}_{k} = \pmb{R}_{k,\mathrm{TX}}^{\frac{1}{2}} \pmb{h}_{\mathrm{iid},k},~~~~~ (k = 1,2,\cdots,K) \nonumber
 \end{equation}
where, $\pmb{h}_{{\mathrm{iid},k}}$ is an $M \times 1$ column vector whose elements follow the independent and identically distributed (i.i.d.) complex zero-mean, unit variance Gaussian random distribution.

 Next, we consider the channel model for the proposed system in which the RF lens is deployed with the antenna array at the BS. Here, the presence of an RF lens enables the signal energy to be focused on a certain region of the antenna array at the BS depending on each user's AoD. We assume that the AoD of the signal is equivalent to the value of the angle location of the $k$-th user. 
 
 Since the location of the energy-compacted region at the ULA is dependant on AoD, we define the normalized power profile vector as 
  \begin{equation}
  \pmb{a}(\theta_{k}) = [a_{k1},\ a_{k2},\cdots\ ,a_{kM}]^{{T}}. \nonumber
  \end{equation}
   The element $a_{{km}}$ denotes the additional power factor captured by the ${m}$-th antenna at the BS. The additional power factor captured by the $m$-th antenna can be modeled as an integral calculus of the continuous power density function $A(\theta)$ \cite{zhang2014lens}. We also define a power profile matrix $\pmb{A} = [\pmb{a}(\theta_{{1}})\ \pmb{a}(\theta_{{2}})\cdots\ \pmb{a}(\theta_{{K}})]^{T}$ of size $K\times M$, where the rows of the matrix are the power profile vectors from all users.
  
    Since the RF lens is a passive device with linear and invertible transfer function, we can therefore simply multiply the additional power profile vector for the $k$-th user to the channel vector $\pmb{h}_k$ without the RF lens. 
By adding the power profile vector, the previous channel model for the ${k}$-th user is modified to
\begin{equation}
{{\tilde{\pmb{h}}}_{{k}}} = \sqrt{\pmb{a}(\theta_{{k}})}\circ {\pmb{h}_{{k}}}, \nonumber
\end{equation}
where $\circ$ is an entry-wise product of two vectors (or matrices). 
The modified channel matrix ${\tilde{\pmb{H}}}$ for all users is expressed as
\begin{equation}
{\tilde{\pmb{H}}} = [\pmb{a}(\theta_{1})|\pmb{a}(\theta_{{2}})\cdots|\pmb{a}(\theta_{{K}})]^{T}\circ {\pmb{H}} = \pmb{A}\circ{\pmb{H}}. \nonumber
\end{equation}  

Let ${\pmb{g}}_{k}$ be the vector-normalized transmit precoding vector for the $k$-th user, ${\pmb{g}}_{k}={\pmb{f}}_{k}/{(\sqrt{K}\norm{{\pmb{f}}_{k}})}$, where vector $\pmb{f}_{k}$ is a column vector of a precoding matrix, $\pmb{F}$ (We consider ZF and MRT precoding in our system). Then, the received signal at the downlink transmission for the ${k}$-th user is expressed as 
\begin{equation}
\begin{split}
{y_{{k}}} &= \sqrt{P_{\mathrm{t}}}{{\tilde{\pmb{h}}}_{{k}}^{T}}\pmb{g}_{{k}}s_{{k}}+\sum_{j=1,j\neq k}^{K‎}\sqrt{P_{\mathrm{t}}}{{\tilde{\pmb{h}}}_{{k}}^{T}}\pmb{g}_{{j}}s_{{j}}+{n_{k}}\\
&= \sqrt{\frac{P_{\mathrm{t}}}{K}}{{\tilde{\pmb{h}}}_{{k}}^{T}}\frac{\pmb{f}_{k}}{\norm{{\pmb{f}}_{k}}}s_{{k}}+\sum_{j=1,j\neq k}^{K‎}\sqrt{\frac{P_{\mathrm{t}}}{K}}{{\tilde{\pmb{h}}}_{{k}}^{T}}\frac{\pmb{f}_{j}}{\norm{{\pmb{f}}_{j}}}s_{{j}}+{n_{k}}, \nonumber
\end{split}
\end{equation}
where ${s}_{k}$ and ${n}_{k}$ denote the transmit symbol at the downlink, and the additive white Gaussian noise vector, respectively, for the $k$-th user and ${P}_\mathrm{t}$ represents the total transmit power from the BS.

\section{Field Distribution of RF Lens}
\label{Field}

The channel matrix of the lens-embedded MIMO system can be varied due to the additional power profile from the different AoDs of the users. It is, therefore, important to accurately estimate the value of additional power gain. Since the signal power can be achieved from the field amplitude of the wave, a method is provided to successfully estimate (or calculate) the field distribution of RF lens. Here, we present a reversed method to analyze the plane wave signal transmitted to the RF lens with a different AoD. For computationally simple and accurate analysis, however, we conduct analysis based on Fourier optics and BPM.
In \cite{zhang2015}, the array response for the lens antenna array at the focal arc with critical antenna spacing is derived to be the sinc function that corresponds to the AoD of the signal. From a lens design perspective, however, not only is it hard to fabricate the negligible thickness planar dielectric lens but it is also hard to match the phase profile calculated. Also, even if it is well designed, the beams are not exactly focused in the focal arc making it hard to match the theoretical beam pattern, potentially giving rise to a discrepancy within the measured data. The BPM, which is based on traditional Fourier optics and Huygens' principle, may calculate all signal power data at any distance and at any coordinates. The alternative solution, BPM analysis here can give insights into the analysis of the lens-embedded massive MIMO system since there are difficulties in performing a full-wave simulation using a commercial EM simulation tool for propagating and focusing beams because of the structural complexity and the feeding assignment of the antenna array.

\subsection{Phase Transform Function of the RF Lens}
\begin{figure}[!t]
	\setlength{\belowcaptionskip}{-15pt}
  \centerline{\resizebox{1.0\columnwidth}{!}{\includegraphics{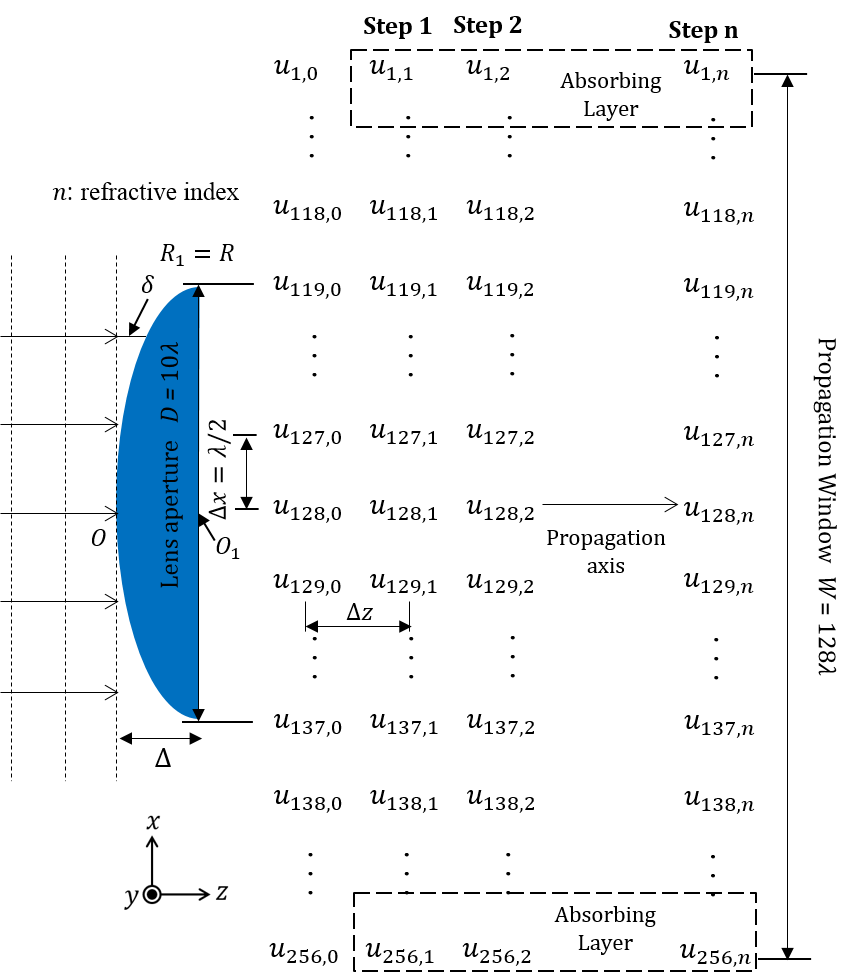}}}
   \caption{BPM matrix model for the focused beam from the lens.}
   \label{Fig:bpmmatrix}
\end{figure}

The field distribution in front of the lens aperture should be firstly calculated to conduct an iterative BPM (See Fig.~\ref{Fig:bpmmatrix}). A thin lens where its aperture is located on the \xyplane{} acts as a phase transformation, if a beam entering at coordinates $(x,y)$ on one face emerges at approximately the same coordinates on the opposite face, that is, if there is negligible transition of the beam inside the lens. Since we consider only 1D array model, 2D analysis is sufficient.
 Let $\Delta$ be the maximum thickness of the lens, and $\delta$, which is a function of $x$ and $y$, be the remaining distance of free space that lies between the lens and the planes tangent to the entrance of the lens. The total phase shift accumulated by the normal incident plane wave can be expressed as,

\begin{equation}
\Phi = \kappa(\underbrace{\delta}_\text{Air}+\underbrace{n(\Delta-\delta)}_\text{Lens}) = \kappa n \Delta-\kappa(n-1)\delta, \nonumber
\end{equation}
where $\kappa = \frac{2\pi}{\lambda_{0}}$.
Within the paraxial approximation, a simple mathematic calculation shows that $\delta$ can be written as \cite{opticbook},

\begin{equation}
\delta(x,y) = {\frac{x^2+y^2}{2}}{\left(\frac{1}{R_1}-\frac{1}{R_2}\right)}.\nonumber
\end{equation}
The phase transform introduced by the lens can therefore be described as
\begin{eqnarray}
\Phi(x,y) &=& \kappa n{\Delta}-\kappa(n-1)\left(\frac{x^2+y^2}{2}\right){\left(\frac{1}{R_1}-\frac{1}{R_2}\right)} \nonumber\\
&\stackrel{(a)}{=}& \kappa n{\Delta} -\frac{\kappa \left(x^2+y^2\right)}{2f},\nonumber
\end{eqnarray}
where equality $(a)$ is given from the lens equation, 
\begin{equation}
\frac{1}{f}=(n-1)\left[\frac{1}{R_1}-\frac{1}{R_2}\right]. \nonumber
\end{equation}
By considering a 2D configuration, the phase shift of RF lens, $\Phi(x;f)$,  is given by
\begin{equation}
\Phi(x;f) = \kappa n{\Delta}-\kappa\frac{x^2}{2f}. \nonumber
\end{equation}
The phase transform function $\Psi(x,f)$ can be re-written as:
\begin{equation}
\Psi(x;f) = \exp\left({-\kappa\frac{x^2}{2f}}\right). \nonumber
\end{equation}

For the oblique incident plane wave where $\theta_\text{ele}\theta_\text{azi}\neq0$, a phase difference occurs as the ray deviates from the ray that is headed to the point of origin.
We assume here that the phase shift by the lens is the same as those of normal incident plane waves, an assumption that may be inaccurate. To calculate the exact value of the phase shift function of the lens, the ray-tracing method should be considered.  Here, we assume the thin lens case, so that the ray from $(x,y,0)$ at the input plane corresponds to $(x,y,\Delta)$ at the output plane.
 This contributes to the total phase shift at the output plane, which is represented by:
\begin{equation}
\begin{split}
\Phi(x,y;f) = & -\kappa\sqrt[\leftroot{-3}\uproot{3}]{(x\sin{\theta_\text{ele}})^2+(y\sin{\theta_\text{azi})^2}} \\
& +\left(\kappa n{\Delta}-\kappa\frac{x^2+y^2}{2f}\right). \nonumber
\end{split}
\end{equation}

\subsection{Assumptions}
Here, we introduce the numerical method that can iteratively calculate the sampled field distribution for every sampling distance $\Delta z$. You can see the basic procedure of BPM in Fig.~\ref{Fig:bpmmatrix}.  We calculate the additional power factor captured by each antenna component with this method. 
To simplify the mathematical analysis and reduce the computational complexity,  we make several assumptions as follows:
\begin{itemize} 
\item RF Lens: We assume a thin lens to simplify the analysis of phase distribution at the lens. The surface geometry is assumed to be rotationally symmetric, and the parabolic approximation is used on the lens geometry.
\item Conservation of the power: The phase function of the lens is utilized in calculating the initial field distribution $\pmb{u}_{0}$ at the lens surface, since the lens is a phase shifter from an EM point of view. By calculating the field distribution at the distance $\ell$, we are able to acquire the fraction of the power captured by each antenna element at the BS. Due to the conservation of power, the sum of all the entries in each power profile vector is always the same.
\item Reciprocity: BPM is used to calculate the field distribution (or power distribution) of propagating and focusing beams at the uplink transmission. Here, we assume that the BS can support users in the downlink transmission by using the same amplitude profiles acquired from the uplink transmission. The amplitude of the signal at the uplink transmission can be distinctively acquired since the spatial power distribution of an incident wave passing through the RF lens is dependent on the angle of arrival (AoA). We can, therefore, use these values as profiles for the downlink. 

\item Scalar representation of a wave equation: To analyze an RF lens antenna using Maxwell's equations, it is unnecessary to consider all the components of the electric and magnetic fields. Since the vector wave equation is obeyed by both the electric field, $\overrightarrow{E}$, and the magnetic field, $\overrightarrow{H}$, all components of these vectors obey an identical scalar wave equation, for $E_x$, (for example, 
$\nabla^2{E_x}-\frac{{n}^2}{{c}^2}\frac{{\partial}^2E_x}{{\partial t}^2}=0$). We, therefore, use $U$ as a scalar field distribution and further analysis will be done on only $U$. To calculate the total intensity of the wave, the squared sum of all corresponding components should be considered. Detailed information regarding this can be found in \cite{opticbook}. 
\end{itemize}

\begin{figure*}[t]
        \begin{subfigure}[b]{1.01\columnwidth}
                \includegraphics[width=0.98\columnwidth,height = 3cm]{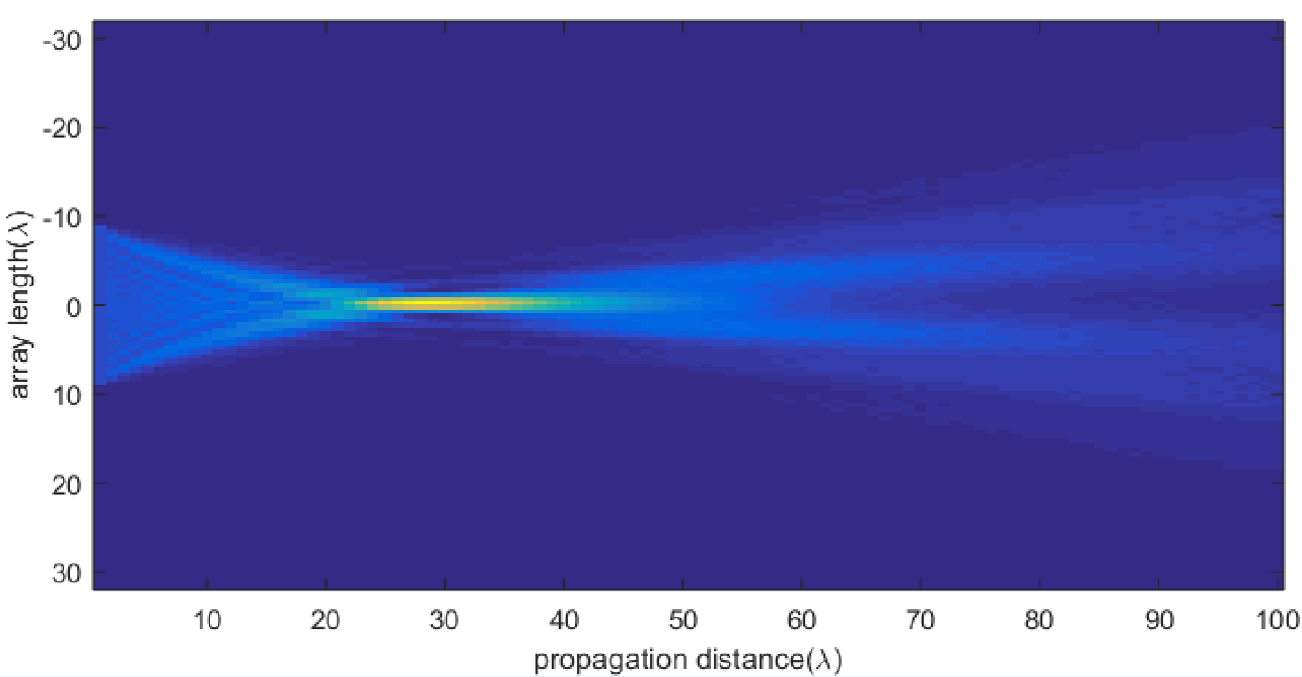}
                \caption{E-field distribution of an RF lens with $\text{AoD} = 0^{\circ}$ and $f = 40\lambda$}
                \label{efield1}
        \end{subfigure}
        \begin{subfigure}[b]{1.05\columnwidth}
                \includegraphics[width=0.98\columnwidth,height = 3cm]{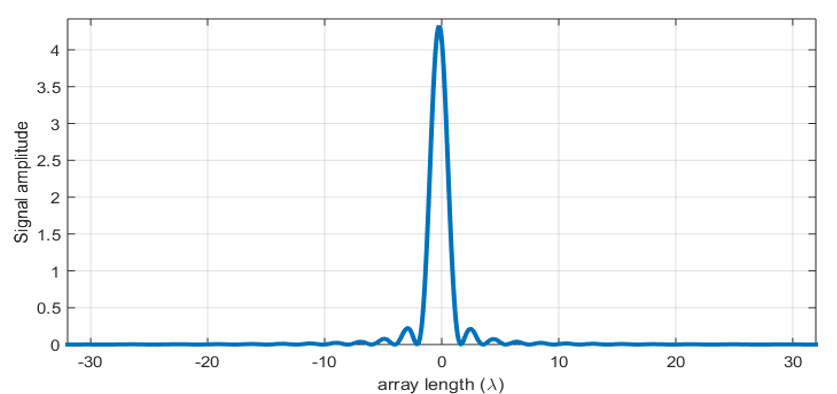}
                \caption{The cross section amplitude at distance $30\lambda$}
                \label{amp1}
        \end{subfigure}
        \begin{subfigure}[b]{1.01\columnwidth}
                \includegraphics[width=0.98\columnwidth,height = 3cm]{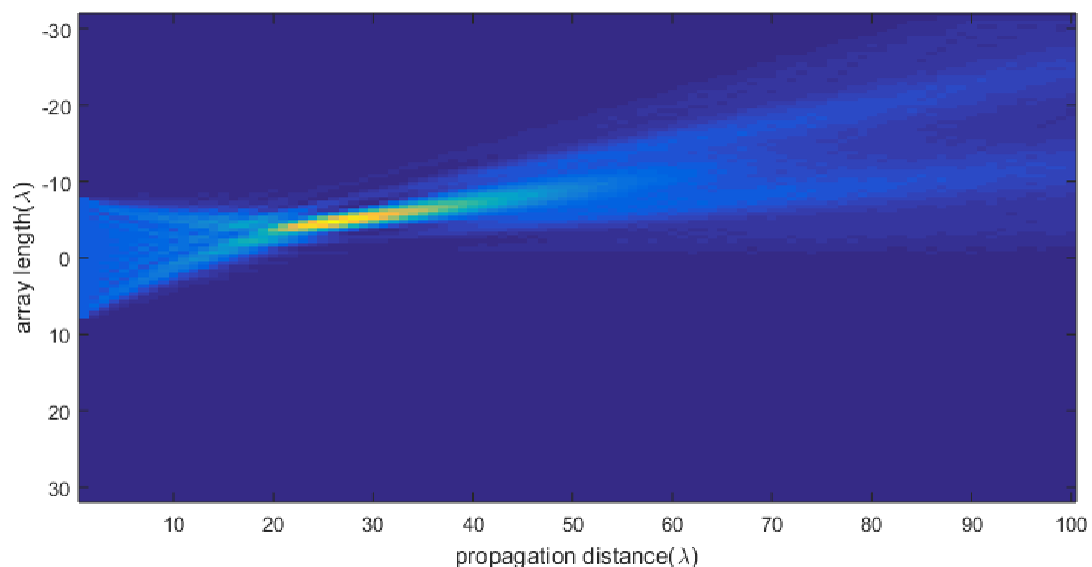}
                \caption{E-field distribution of an RF lens with $\text{AoD} = 15^{\circ}$ and $f = 40\lambda$}
                \label{efield2}
        \end{subfigure}
        \begin{subfigure}[b]{1.05\columnwidth}
                \includegraphics[width=0.98\columnwidth,height = 3cm]{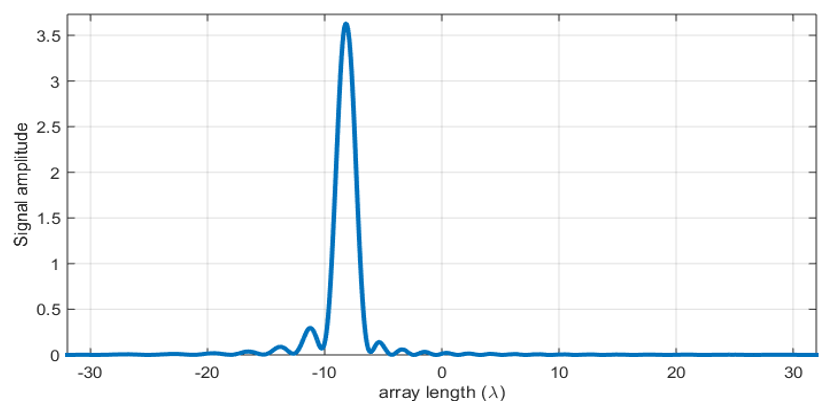}
                \caption{The cross section amplitude at distance $30\lambda$}
                \label{amp2}
        \end{subfigure}
        \setlength{\belowcaptionskip}{-12pt}
        \caption{Propagating and focusing property of incident wave to the lens.}
        \label{Fig:focusing}
\end{figure*}

\subsection{Beam Propagation Method}
Assume that the lens aperture is on the $x$-$y$ plane, and the beam is being propagating in the direction of the positive $z$-axis.  From Fourier optics, the propagations of EM waves through the RF lens can be expressed as follows \cite{opticbook}:

\begin{equation}
{U}{(x,y)} 
= \frac{{e}^{{jkz}}}{j\lambda z}{\mathscr{F}}^{-1}{\left[{\mathscr{F}}\left\{{{U}\left(x',y'\right)}\right\}\circ{\mathscr{F}}\left\{{e}^{j\frac{k\left(x'^2+y'^2\right)}{2z}}\right\} \right]}, 
\end{equation}
where $\mathscr{F}$ denotes the Fourier transform operator, ${{U}\left(x',y'\right)}$ is the field distribution at the aperture, ${{U}(x,y)}$ is the field diffraction pattern at distance $z$, and ${{e}^{jk\left(x'^2+y'^2\right)/(2z)}}$ is the quadratic phase factor. Several conditions should be hold such as the observer is said to be in the near field of the aperture, the surface geometry is assumed to be rotationally symmetric, and the focal points are placed on its axis. Off-axis aberrations should also be ignored. Details of derivation is in \cite{opticbook}.

 Since our only consideration on array configuration is 1D ULA, we can reduce the dimension to 2D lens model for an amenable analysis. By omitting the additional dimension variables, equation (1) is reduced to the following,
 
 \begin{equation}
{U}{(x)} 
= \frac{{e}^{{jkz}}}{j\lambda z}{\mathscr{F}}^{-1}{\left[{\mathscr{F}}\left\{{{U}\left(x'\right)}\right\}\circ{\mathscr{F}}\left\{{e}^{j\frac{k}{2z}x'^2}\right\} \right]}. 
\end{equation}

The physical meaning of equation (2) is that the field distribution at distance $z$ is obtained by the inverse Fourier transform of the product of two Fourier transforms of the complex field just to the right of the aperture and a quadratic phase exponential. 

 When this equation is found to be valid, the observer is said to be in the region of Fresnel diffraction, or equivalently in the near field of the aperture. Therefore, one may think of setting the propagation step at a relatively small distance, repeatedly resetting the aperture distribution to calculate the accurate field distribution form which can yield the iterative algorithm.

 BPM, in this context, is used to calculate the field distribution (or power distribution) of propagating and focusing beams at the uplink transmission \cite{Roey1981bpm}. To obtain the power profile vector, the discrete field distribution is sampled in the $x$-axis. The sampling distance is given as $\Delta x$, where the range of the sampling is constrained to propagation window, $W$. We decide $W$ as multiples of $\Delta x$. In Fig. 3, the first column vector is the field distribution just to the right of the lens aperture, which is the vector with $N_s = W/\Delta x$ elements. The same topology is adapted for $\pmb{u}_n$, which is a power profile vector at the distance $n\Delta z$, where $\Delta z$ denotes the sampling distance on the $z$-axis.
\begin{equation}
\pmb{u}_n={U}_{z = n \Delta z}{(x=m\Delta x, m = -N_s/2,\cdots, N_s/2)}. \nonumber
\end{equation}
 The phase transform function of the lens is utilized in calculating the initial field distribution $\pmb{u}_{0}$ at the lens surface, since the lens,  from an EM point of view, is a phase shifter. $\pmb{u}_{0}$ can be obtained from the phase transform function,
\begin{equation}
\pmb{u}_0 = \Psi(x;f) = \exp\left({-\kappa\frac{{\left(m\Delta x\right)}^2}{2f}}\right). \nonumber
\end{equation}

The Fourier transform of the current field distribution is multiplied by the wave system function $\pmb{h}_\mathrm{sys}$,
\begin{eqnarray}
\label{sys_eqn}
&\pmb{u}_{n}& = \frac{e^{jk\Delta z}}{j\lambda \Delta z}{\mathscr{F}}^{-1}{\big[{\mathscr{F}}\{\pmb{u}_{n-1}\}\circ \pmb{h}_\mathrm{sys} \big]},  \\  
&\pmb{h}_\mathrm{sys}& = {{\mathscr{F}}}\left\{e^{j\frac{k(m\Delta x)^2}{2\Delta z}}\right\}. \nonumber
\end{eqnarray}

Thus, if the initial field distribution at the lens surface is given as $\pmb{u}_{0}$, we can inductively calculate the next step of discrete field distribution. The proportionality of power density to the squared magnitude of the \pmb{u} vector leads us to define the intensity of a wave as the squared magnitude of the $\pmb{u}_{n}$.

\begin{equation}
\pmb{p}_{n}=c|\pmb{u}_{n}|^2, \sum_{i=0}^{i=N_s}p_{ni}=M. \nonumber
\end{equation}

 Note that this intensity is not identical to the power density, but directly proportional. According to the conservation of power, the sum of all the entries in each power profile vector is always the same. By adding all the components within the range of the antenna, we are able to acquire the fraction of the power captured by each antenna element at the BS. (For continuous power density function, an integral is needed to calculate the fraction of the power.)
We can therefore estimate the normalized power profile vector at the $z = n\Delta z$ with the angle of $\theta_{k}$, $\pmb{a}(\theta_{k}, z = n\Delta z)= [a_{k1},\ a_{k2},\cdots\ ,a_{kM}]^{{T}}$, as follows
  \begin{equation}
 	 a_{{km}} = c{{‎‎\sum}}_{i=(m-1)\lfloor{\frac{DNs}{WM}}\rfloor+1}^{i=m\lfloor{\frac{DNs}{WM}}\rfloor}p_{ni}. \nonumber
  \end{equation}
   The element $a_{{km}}$ denotes the additional power factor captured by the ${m}$-th antenna at the BS. 
 

\begin{table}[h]
\centering
\begin{tabular}{| c || c | c | c | c |} \hline G. Focal Length & 20$\lambda$ & 30$\lambda$ & 40$\lambda$ & 50$\lambda$\\ \hline \hline Peak Distance & 16$\lambda$ & 20$\lambda$ & 27$\lambda$ & 33$\lambda$\\ \hline  
Focusing Intensity & 6.11x &  4.92x &  3.47x &  2.88x\\ \hline  
\end{tabular} 
\caption{Beam propagation of focused beams with geometrical focal lengths of $20\lambda$, $30\lambda$, $40\lambda$, and $50\lambda$.}
\label{cal1}
\end{table}

Figure~\ref{Fig:focusing} illustrates, using the BPM algorithm, the propagating and focusing properties of a wave through the RF lens.
For the RF lens with $20\lambda$, $30\lambda$, $40\lambda$ and $50\lambda$ of geometrical focal length, the beam width, peak distance, and field intensity are calculated in Table~\ref{cal1}. Starting from the aperture of the lens, which is given as $20\lambda$, an iterative propagation method is adapted where the sampling distance $\Delta x$ and $\Delta z$ are $1\lambda$ and $1\lambda$. One can observe the difference between the geometric focal length and peak distance, which corresponds well with diffraction theory \cite{hansen1985}. The field intensity of the wave is intensified by 6.11, 4.92, 3.57 and 2.88 at the peak distance of the wave.\footnote{In practice, the radiated beam from a horn antenna is calculated accurately with the modified multimode Gaussian beam and transferred through the lens using ray tracing. Also, the previous discussion of an RF lens and BPM considered a thin lens approximation. Since we attempt to find an accurate data, for future work, thick lens analysis will be carried out with more practical setup.}

\section{Proposed Channel Feedback Algorithm}
\label{algorithm}
Now we consider an FDD operation with limited feedback using a codebook-based channel quantization. The users only need to feed-back their channel direction information (CDI) to the BS to minimize the channel feedback overhead. A quantization codebook, therefore, consists of $\mathcal{N}M$-dimensional unit norm vectors ${\pmb{C}}$ = $[\pmb{c}_{1}|\pmb{c}_{2} ...|\pmb{c}_\mathcal{N}]$, where $\mathcal{N}$ is its cardinality (a positive integer)  and ${\pmb{C}}$ is known to be at both the BS and users. In the finite-rate feedback model where each receiver quantizes its channel to $B$ bits and feeds-back the bits perfectly to the BS, $\mathcal{N}$ is equal to $2^B$. The beamforming vector that maximizes the inner product with the actual channel direction vector will be chosen from Codebook ${\pmb{C}}$. Thus, the codebook index ${j_{{k}}}$, selected for the ${k}$-th user, can be acquired from the formula: ${j_{{k}}} = {\mathrm{arg}_{{j}}}\mathrm{max}|{\pmb{h}_{k}^{\ast}}\pmb{c}_{j}|$. In practical systems, the users only need to feed-back an index of the codebook vector ${j_{{k}}}$ that correlates the best with their channel vector. Here we briefly introduce RVQ, which has been widely used for performance analysis.

\subsection{Random Vector Quantization (RVQ)}
\label{RVQ}
One of the easiest and most adaptable ways to design a codebook is to generate one randomly. In RVQ beamforming with the feedback bits ${B}$, a codebook is generated as $\pmb{W} = [\pmb{w}_{1}|\pmb{w}_{2}\cdots |\pmb{w}_{2^B}]$ of size $M\times2^B$. The column vector $\pmb{w}_{j}$ $(j = 1,2,\cdots,2^B)$ denotes the codebook vector. Note that each column, $\pmb{w}_{j} = [{w_{i1}} {w_{i2}} \cdots {w_{iM}}]^{T}$, is an i.i.d. isotropic vector on the $M$-dimensional unit sphere, as are the channel directions. In a spatially-correlated channel, the RVQ codebook, $\pmb{W}_{\mathrm{rvq}}$, is given as, $\pmb{W}_{\mathrm{rvq}} = ({\pmb{W}^T}{\pmb{R}^\frac{1}{2}_{\mathrm{TX}}})^{T} = [\pmb{w}'_{1}|\pmb{w}'_{2} \cdots |\pmb{w}'_{2^B}]\in\mathbb{C}^{M\times2^B}$. If a codebook index is chosen as ${j}$, the estimated channel at the BS is selected as $\hat{\pmb{h}}_{{k}}$ = $\pmb{w}'_{j}$.

\smallskip
\subsection{Multi-Variance Codebook Quantization (MVCQ)}
 Here, we propose a new feedback algorithm based on RVQ by considering the power profile vectors of the users. Under the presence of the RF lens within the ULA at the BS, the conventional RVQ cannot reflect the changed channel characteristics of different variances. Since our system model assumes that the users are forward-fed the information of the angle where they are located in a cell, they can exploit the power profile vectors in their codebook designs. The channel vectors associated with the antennas at the BS have different variances since the energy is focused on a subset of antennas at the ULA after passing through the RF lens. We can, therefore, simply multiply the $m$-th power profile element by the $m$-th element of the conventional RVQ codebook vector so that the variance of each entry in the constructed codebook vector will shift to the desired level. Thus, the entries of the new codebook vector for the $k$-th user are made to have different variances, $a_{{m}}(\theta_{{k}})$, and we call this new channel quantization method, MVCQ. Thus, the proposed codebook $\pmb{W}_{\mathrm{mvcq},k}$ for the $k$-th user is expressed as  
\begin{equation}
\begin{split}
\pmb{W}_{\mathrm{mvcq},k} &= \left[\sqrt{\pmb{a}(\theta_{{k}})}\circ\pmb{w}'_{1}| \sqrt{\pmb{a}(\theta_{{k}})}\circ\pmb{w}'_{2} \cdots |\sqrt{\pmb{a}(\theta_{{k}})}\circ\pmb{w}'_{2^B}\right]\\
 &= [\pmb{w}''_{1}|\pmb{w}''_{2}\cdots|\pmb{w}''_{2^B}]\nonumber,
\end{split}
\end{equation}
where $\pmb{w}''_{j}$, the column vector of $\pmb{W}_{\mathrm{mvcq},k}$ follows the distribution of, $\pmb{w}''_{j} = [{w}''_{i1} {w}''_{i2} \cdots {w}''_{iM}], {w}''_{im}\sim \mathcal{C}\mathcal{N}(0,a_{{m}}(\theta_{{k}}))$. 


We can analyze, roughly, the effect of MVCQ in terms of the power correlation matrix and signal-to-noise-plus-interference ratio ($\text{SINR}$). The power correlation matrix $\pmb{\Psi}$ is defined as $\pmb{\Psi} = \sqrt{\pmb{A}^T}\sqrt{\pmb{A}}\nonumber$, where $\pmb{A}$ is the power profile matrix. The power correlation matrices for the system without the RF lens ($\pmb{\Psi}^{(1)}$) and with the RF lens ($\pmb{\Psi}^{(2)}$) are given, respectively, as,
\begin{equation}
\begin{split}
\pmb{\Psi}^{(1)} =
 \begin{pmatrix}
  1 & 1 & \cdots & 1 \\
  1 & 1 & \cdots & 1 \\
  \vdots  & \vdots  & \ddots & \vdots  \\
  1 & 1 & \cdots & 1
 \end{pmatrix},
\smallskip
{\pmb{\Psi}}^{(2)} =
 \begin{pmatrix}
  1 & r_{1,2} & \cdots & r_{1,K} \\
  r_{2,1} & 1 & \cdots & r_{2,K} \\
  \vdots  & \vdots  & \ddots & \vdots  \\
  r_{K,1} & r_{K,2} & \cdots & 1,
 \end{pmatrix}&,\nonumber\\
\end{split}
\end{equation}
where $r_{j,k}=r_{k,j}=\sqrt{\pmb{a}(\theta_{{j}})^{T}} \sqrt{\pmb{a}(\theta_{{k}})}/{M}$.
Since  ${\pmb{a}(\theta_{k})}/{M}$ is a normalized power profile for the $k$-th user, $r_{j,k} = r_{k,j} =$ $\sqrt{\pmb{a}(\theta_{{j}})^{T}} \sqrt{\pmb{a}(\theta_{k})}/M <1, \forall j\neq k$ and $r_{k,k} = 1.$ 

Also, the $\text{SINR}$ for the $k$-th user is approximated as
\begin{equation}
 \text{SINR}_{k} \simeq \frac{P_{t}/{K}\pmb{\Psi}^{(i)}_{k,k}\left|{\pmb{h}}_{k}^{T}{\pmb{f}}_{k}/{\norm{{\pmb{f}}_{k}}}\right|^2}{P_{t}/{K}\sum_{j=1,j\neq k}^{K‎}\pmb{\Psi}^{(i)}_{k,j}\left|{\pmb{h}}_{k}^{T}{\pmb{f}}_{j}/{\norm{{\pmb{f}}_{j}}}\right|^2+1},\nonumber
\end{equation}
where $i=1$, or $2$ for the system without or with the RF lens, respectively.

While all the elements in $\pmb{\Psi}^{(1)}$ are 1, $\pmb{\Psi}^{(i)}_{k,j}, k\neq j$, which denotes the off-diagonal term of ${\pmb{\Psi}}^{(2)}$, they can be decreased to less than 1. This implies that the desired signal power term is the same for both cases while the interference power term of SINR is greatly decreased in the RF lens-embedded massive MIMO systems. 
 We can, therefore, insist that MVCQ improves the SINR for those users who are sufficiently resolvable at the BS.


 In practice, it is possible to generate the codebook based on the measured/estimated AoD of each user, but at the mobile stations (MSs) its computation complexity could be burdensome. To resolve this issue, we could first generate the codebook for every antenna type and piece of AoD information. Then, the BS and the MS just need to share the control information such as codebook index through limited feedback and feedforward.



\subsection{Low-Complexity Estimation Methods}

The performance of an MVCQ critically depends on the estimation accuracy of power distribution, which corresponds to the additional coefficients $\pmb{a}(\theta_k)$. 
We therefore introduce a power distribution function or sub-algorithm that can be simply adapted to the MVCQ. By coarsely approximating the power distribution function as a well-known distribution, like a Gaussian function, the system has to decide only a few parameters to estimate the power distribution.

\subsubsection{Gaussian}     
The normalized Gaussian power distribution is adapted~\cite{2013gaussian}. To specify the function, three parameters should be determined. The power coefficient for Gaussian MVCQ is given as
\begin{equation}
\pmb{a}(y;\theta)=p(\theta)\exp{\left(-\left(\frac{y-q(\theta)}{r(\theta)}\right)^2\right)}. \nonumber
\end{equation} 
For the RF lens where the focal length is given as $40\lambda$ and the electric permittivity of the lens is set to 2.4, we have tested the 1-Gaussian model for the fixed distance of $\ell = 30\lambda$.

Table~II shows the estimated coefficient of the 1-Gaussian fitting model by changing the incident angle of the signal. Three parameters, $p(\theta)$, $q(\theta)$, and $r(\theta)$, are provided for eight-user cases with different incident angles. By adapting the curve fitting function, we show that each parameter can be formulated as

\begin{displaymath}
   \left\{
     \begin{array}{lr}
       p(\theta) = 2.151\exp{\left(-(\frac{\theta+0.08102}{4.075})\right)}, \\
       \smallskip
       q(\theta) = -0.9714\theta-0.01643, \\
       \smallskip
       r(\theta) = 4.644\exp{\left(-(\frac{\theta-927.7}{528.6})\right)}. \\
     \end{array}
   \right.
\end{displaymath} 
If we also consider the distance between the lens and the antenna array, $\ell$, for fitting, each coefficient has to be modeled as a 2D function. For simplicity, we assume in this paper, a fixed distance. 

\begin{table}[!t]
\centering
\begin{tabular}{|c||c|c|c|c|c|c|c|} 
\hline $\theta$ & \small $-15^{\circ}$ & \small $-10^{\circ}$ & \small $-5^{\circ}$ & \small $0^{\circ}$ & \small $5^{\circ}$ & \small $10^{\circ}$ & \small $15^{\circ}$\\ 
\hline
\hline $p(\theta)$ & \small 1.81 & \small 2.01 & \small 2.12 & \small 2.14 & \small 2.10 & \small 1.97 & \small 1.73\\ 
\hline $q(\theta)$ & \small 1.66 & \small 1.10 & \small 0.54 & \small -0.01 & \small -0.57 & \small -1.13 & \small -1.70\\  
\hline $r(\theta)$ & \small 0.22 & \small 0.21 & \small 0.21 & \small 0.20 & \small 0.21 & \small 0.22 & \small 0.23\\ \hline 
\end{tabular} 
\label{coeff}
\caption{Estimated coefficients of the 1-Gaussian fitting model.}
\end{table}

\begin{figure}[!t]
	\setlength{\belowcaptionskip}{-15pt}
  \centerline{\resizebox{1.0\columnwidth}{!}{\includegraphics{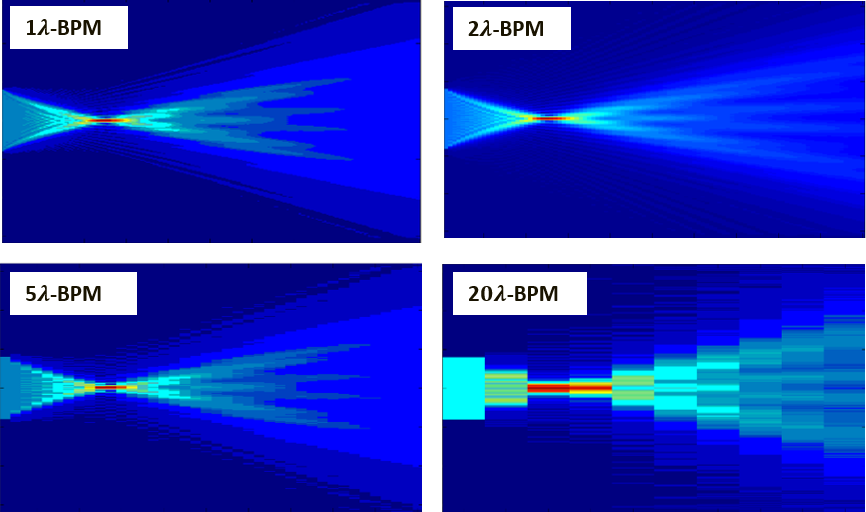}}}
   \caption{Sub-BPM.}
   \label{Fig:subBPM}
\end{figure}

\subsubsection{Sub-BPM}  
Since BPM is an iterative propagation algorithm, the sampling can be done more sparsely to estimate the power distribution. From (\ref{sys_eqn}), the sampling distance $\Delta z$ can be modified freely. Figure~\ref{Fig:subBPM} shows the BPM simulation results of changing the sampling distance by $1\lambda$, $2\lambda$, $5\lambda$, and $20\lambda$. Simulation parameters are the same as the parameters in Fig.~\ref{Fig:focusing}(a). As the sampling distance increases, the estimation accuracy wanes and the complexity of the algorithm decreases.

\begin{figure*}[!t]
	\setlength{\belowcaptionskip}{-15pt}
  \centerline{\resizebox{2.0\columnwidth}{!}{\includegraphics{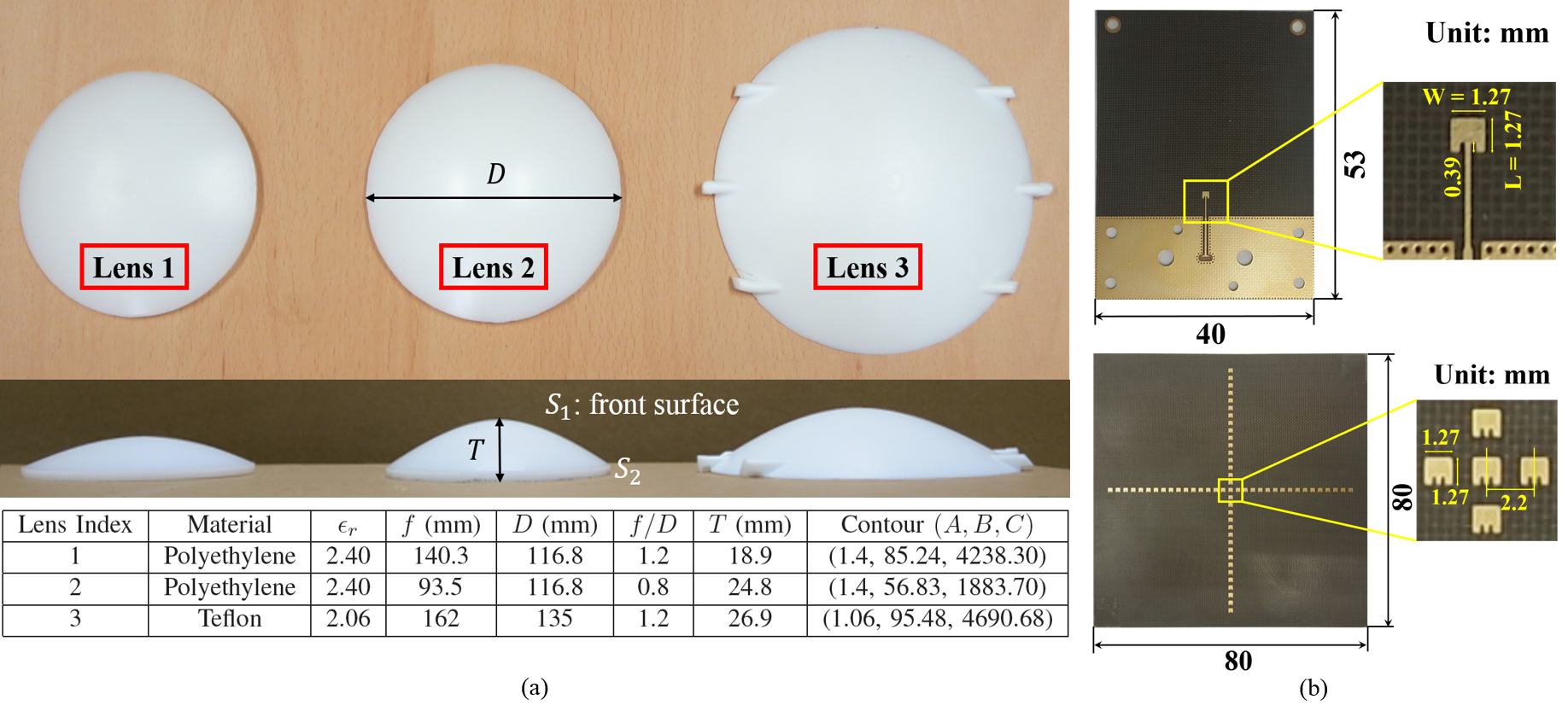}}}
   \caption{(a) Fabricated dielectric lenses and (b) configuration of a single patch probe antenna and array layout (33$\times$33).}
   \label{Fig:Fab}
   \vspace{0.5cm}
\end{figure*}

\section{Measurement System}
To investigate the accordance between the measurements of the power in front of the RF lens and the BPM result, a field measurement system was constructed. Since the performance of a codebook depends significantly on the estimation of the power distribution shape of the RF lens, a more accurate characterization for the power distribution function of the RF lens is needed. The measurements were carried out at 77~GHz; this is the frequency referred to in the following discussions and all the results. The measurement results were compared with the results obtained by BPM. A 77~GHz patch antenna was designed and fabricated as shown in Fig.~\ref{Fig:Fab}(b); for the lens-antenna integration. Two PCB layouts were designed: a single patch antenna that was used for the probe antenna and a 65-cross-aligned patch antenna array. The latter was deployed as an actual BS antenna design. Details of the system components are explained below.




\subsection{RF Lens}
The design principles of lenses are very well known. Geometric optics must be considered while ignoring secondary effects like edge diffraction, surface, or a radiating element impedance mismatch. Here, we consider lens curvature as one of the simplest cases, a hyperbolic lens with dielectric material.

 The dielectric constant of the lens is $\epsilon_r$, which is the square of the refractive index, $n$. Conditions imposed on a dielectric lens are the electrical path length constraint and Snell's law~\cite{antennabook2}. The contours of a few classical lenses (including the hyperbolic lens) can be described with a simple analytic formula. (See Fig. 6) In the simplest case, which is a hyperbolic lens with a flat surface on $S_2$, the contour of $S_1$ is given by 
 \begin{equation}
y_1 = \left[(n^2-1)(x_1-f)^2+2(n-1)(x_1-f)f\right]^\frac{1}{2}. \nonumber
\end{equation}
This can be reformulated as below.

\begin{equation}
y_1^2 = \underbrace{(n^2-1)}_\text{A}\bigg(x_1-\underbrace{\frac{n}{n+1}f}_\text{B}\bigg)^2-\underbrace{\frac{n-1}{n+1}f^2}_\text{C}. \nonumber
\end{equation}
If the values of focal length and the diameter of the lens are fixed, a certain thickness value must be included to fulfill the focusing property of the lens, which is given by
\begin{equation}
T = \frac{1}{n+1}\left\{\left[f^2+\frac{(n+1)D^2}{4(n-1)}\right]^\frac{1}{2}-f \right\}. \nonumber
\end{equation}

Based on this theoretical background, several dielectric lens designs were considered. For the lens material, the study used Teflon ($\epsilon_r=2.06$) and polyethylene ($\epsilon_r=2.40$). The focal length, aperture, thickness, $f/D$ ratio and curvature parameters $A, B, C$ were calculated according to Fig.~\ref{Fig:Fab}(a).

\begin{figure}[!t]
	\setlength{\belowcaptionskip}{-15pt}
  \centerline{\resizebox{1.0\columnwidth}{!}{\includegraphics{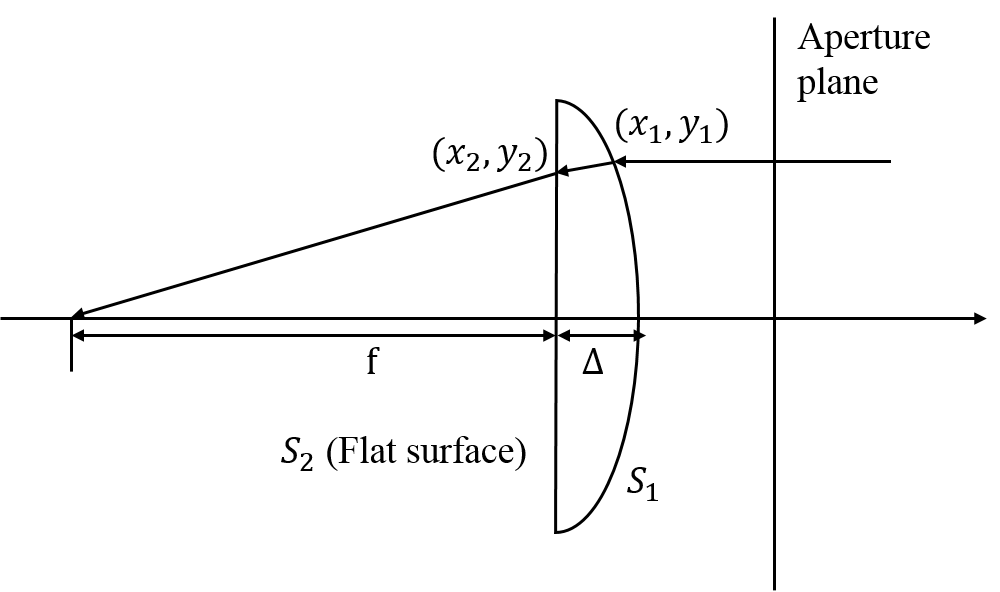}}}
   \caption{Theoretical lens design method.}
   \label{Fig:theo}
\end{figure}

\subsection{Experimental Setup}
\subsubsection{Transmitter}
Used at the transmitter was a horn antenna operating at 70--80~GHz. The power from an oscillator, tunable from 70--80~GHz, was fed through a 0--50~dB calibrated attenuator to the transmitting horn. A reflector was used to lengthen the distance between the transmitter and receiver. 

\smallskip

\subsubsection{Receiver}
For proper antenna design, desirable elements include small size, low directivity, and fairly equal $E-$ and $H-$ plane amplitude patterns. The antenna component was fabricated on 0.127~mm thick Taconic TLY-5 substrate ($\epsilon_r=2.18$, loss tangent $= 0.0075$). For the array design, the dimensions were $W = 1.27~\text{mm}$, $L = 1.27~\text{mm}$. The fabricated patch antennas were aligned in a cross-section where 65 patch antennas ($33\times33$) were used. Array spacing between the patches was set at 2.2~mm, which is $0.56\lambda$ at 77~GHz. During all the measurements, an absorber was placed at the back of the antenna and the lens antenna. In the measurements, a single patch probe antenna was attached at the antenna mounting part. 
A Styrofoam jig, with absorbers attached to the surface, was designed to fix the location of the fabricated RF lens. The center of the lens was aligned with the transmitter and the receiver. 

\subsection{Measurement Procedures}
The measurement was held in an anechoic chamber. To calibrate the measurement system, a horn antenna was used with standard gain at 77~GHz. To determine the spatial power profiles of the propagating beams toward the lens, the probe antenna needed to be placed at different positions along $x$ ($H$-plane) and $y$ ($E$-plane). Considering the alignment problem of the chamber, however, rather than moving the probe antenna, experimenters moved the Styrofoam jig back and forth, as shown in Fig.~\ref{Fig:setting}. The radiation pattern of the receiver antenna was also measured by rotating the antenna in the range of \mbox{[-15$^\circ$,~15$^\circ$]} in the azimuth and elevation angle domains. 
Three types of fabricated lenses in Table II were used. For each configuration, measurements were taken three times by using three sample probe antennas. Presented below are the details of the procedures. 

\begin{figure}[!t]
	\setlength{\belowcaptionskip}{-15pt}
  \centerline{\resizebox{1.0\columnwidth}{!}{\includegraphics{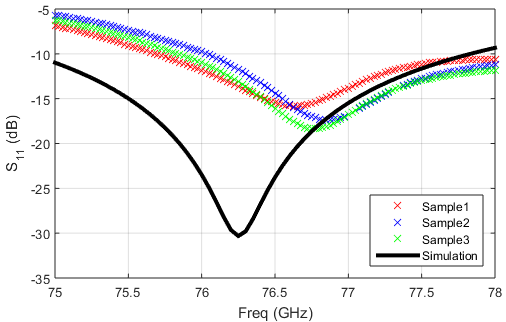}}}
   \caption{Simulation and measured $S_{11}$ parameter data of a single patch probe antenna.}
   \label{Fig:S11}
\end{figure}

\begin{enumerate}
  \item First, the measurements were conducted without an RF lens. Calibration was done initially and $S_{11}$ parameters were measured for each antenna sample. The radiation pattern of the receiver antenna was measured for the range of \mbox{[-90$^\circ$,~90$^\circ$]} in both the elevation and azimuth angle domains. The radiation pattern of the single patch probe antenna without the RF lens was also measured to identify the pure receiver antenna gain.

  \item Installing the RF lens-attached Styrofoam jig in front of the antenna mounting part allowed us to consider the effect of the RF lens. Each component was covered with an absorbing material to minimize any external factors that might affect the field distribution pattern. The distance from the lens was changed by 20~mm steps starting at 40~mm. Azimuth and elevation radiation patterns were measured for the range of \mbox{[-15$^\circ$,~15$^\circ$]} in both the azimuth and elevation angle domains. The axial power distribution was measured on the assumed axis of the beam. 
  
  \item The procedures above were repeated for all three fabricated lenses and the three different probe antenna samples. In total, 90 measurement configurations ($3\times 3 \times 10$) were created to learn the results. 
\end{enumerate}

 \begin{figure*}[!t]
	\setlength{\belowcaptionskip}{-15pt}
  \centerline{\resizebox{2.0\columnwidth}{!}{\includegraphics[width=2.0\columnwidth,height = 8cm]{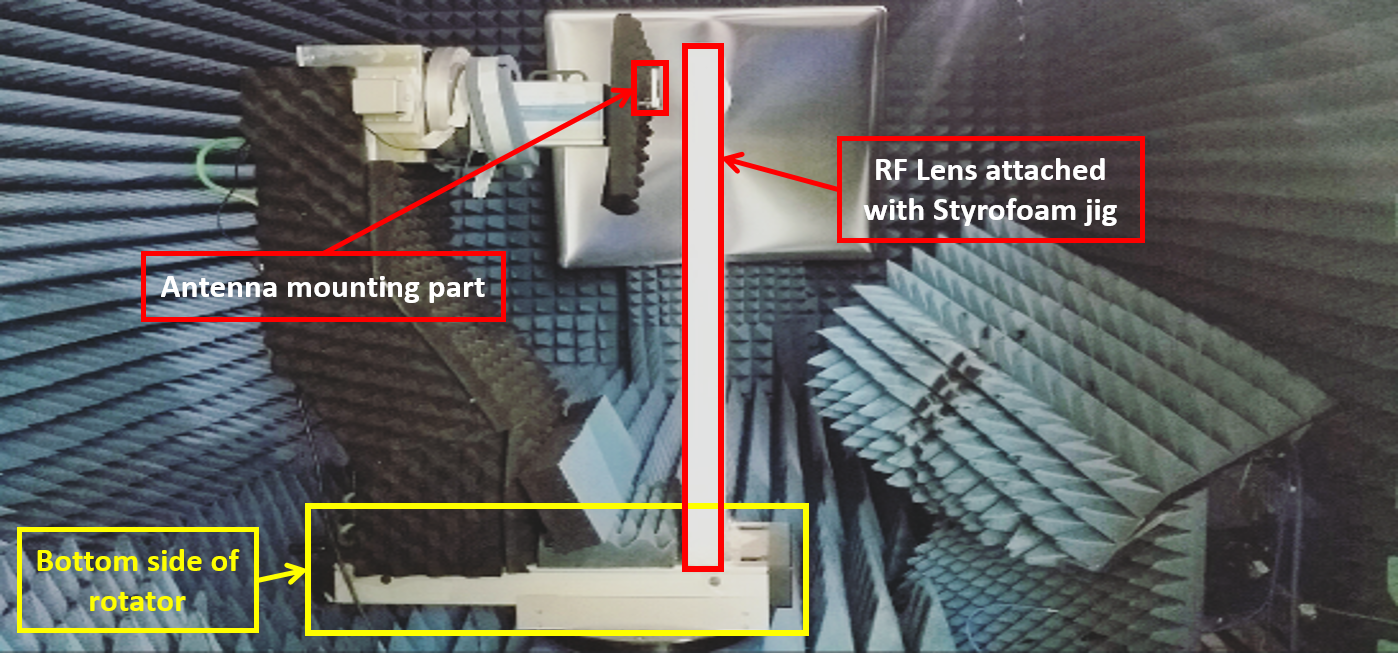}}}
   \caption{Characterization and measurement setup.}
   \label{Fig:setting}
\end{figure*}

\begin{figure*}[!t]
        \centering
        \begin{subfigure}[b]{1.0\columnwidth}
                \centering
                \includegraphics[width=1.0\columnwidth,height = 4.5cm]{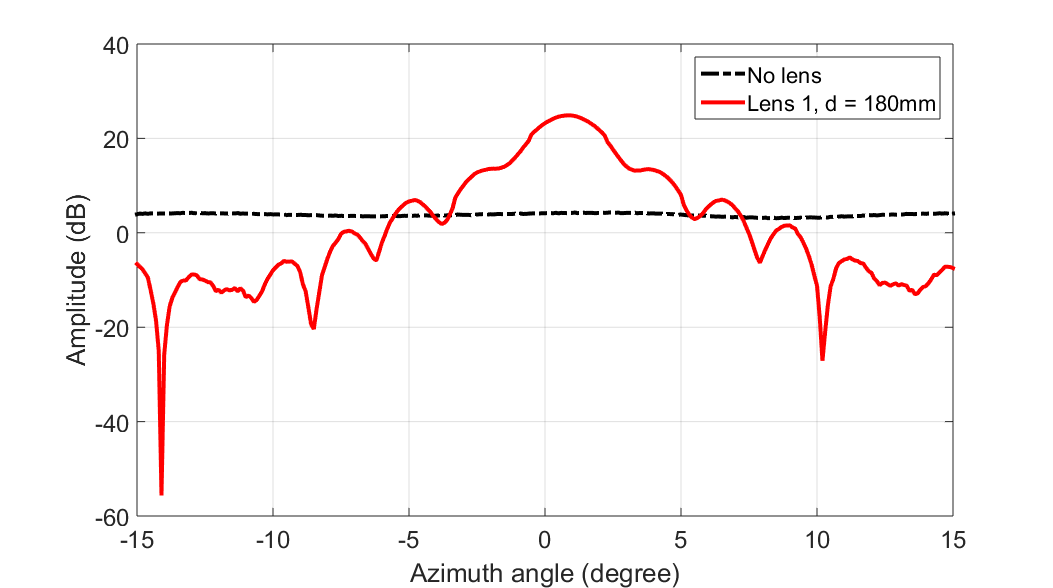}
                \caption{Azimuth radiation pattern with Lens 1}
                \label{azi1}
        \end{subfigure}
        \begin{subfigure}[b]{1.0\columnwidth}
                \centering
                \includegraphics[width=1.0\columnwidth,height = 4.5cm]{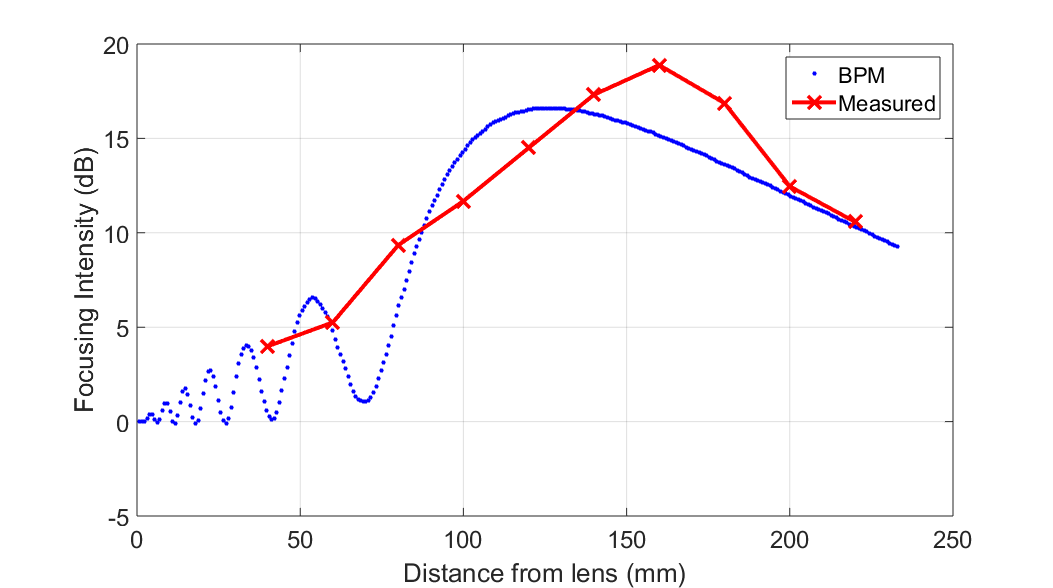}
                \caption{Axial power distribution (focusing intensity) of Lens 1}
                \label{ax1}
        \end{subfigure}
        \begin{subfigure}[b]{1.0\columnwidth}
                \centering
                \includegraphics[width=1.0\columnwidth,height = 4.5cm]{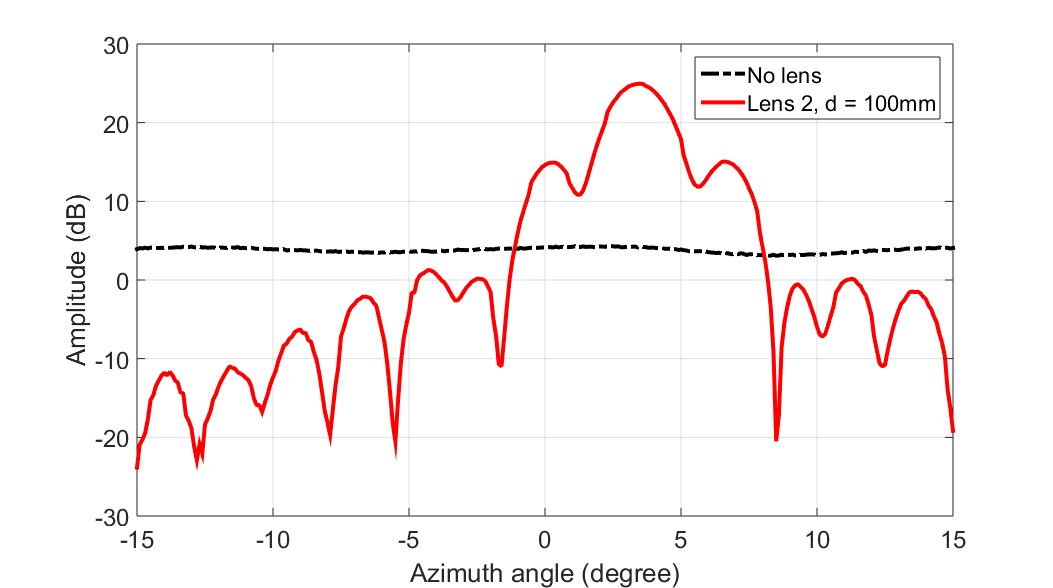}
                \caption{Azimuth radiation pattern with Lens 2}
                \label{azi2}
        \end{subfigure}
        \begin{subfigure}[b]{1.0\columnwidth}
                \centering
                \includegraphics[width=1.0\columnwidth,height = 4.5cm]{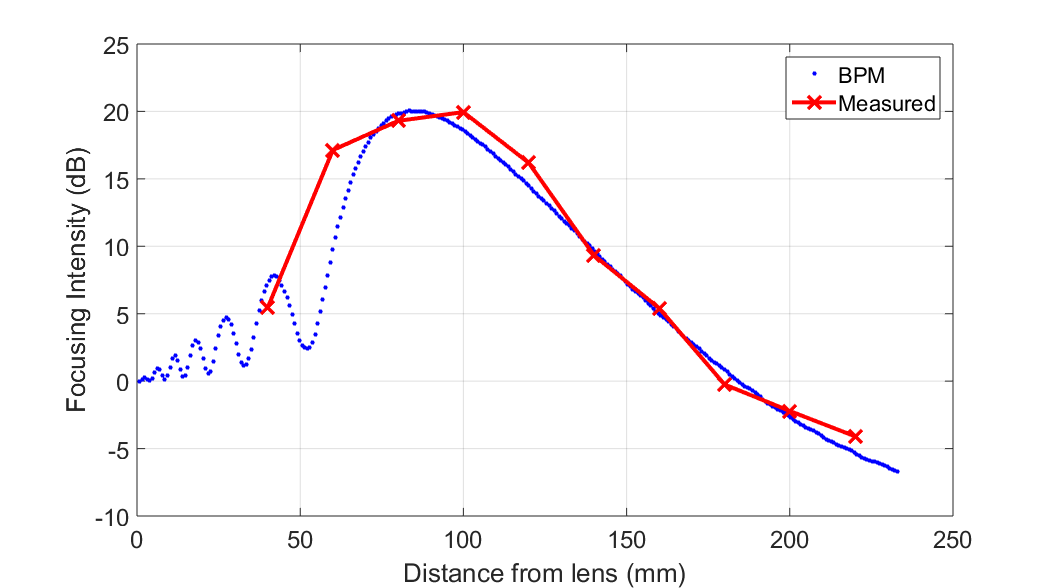}
                \caption{Axial power distribution (focusing intensity) of Lens 2}
                \label{ax2}
        \end{subfigure}
        \begin{subfigure}[b]{1.0\columnwidth}
                \centering
                \includegraphics[width=1.0\columnwidth,height = 4.5cm]{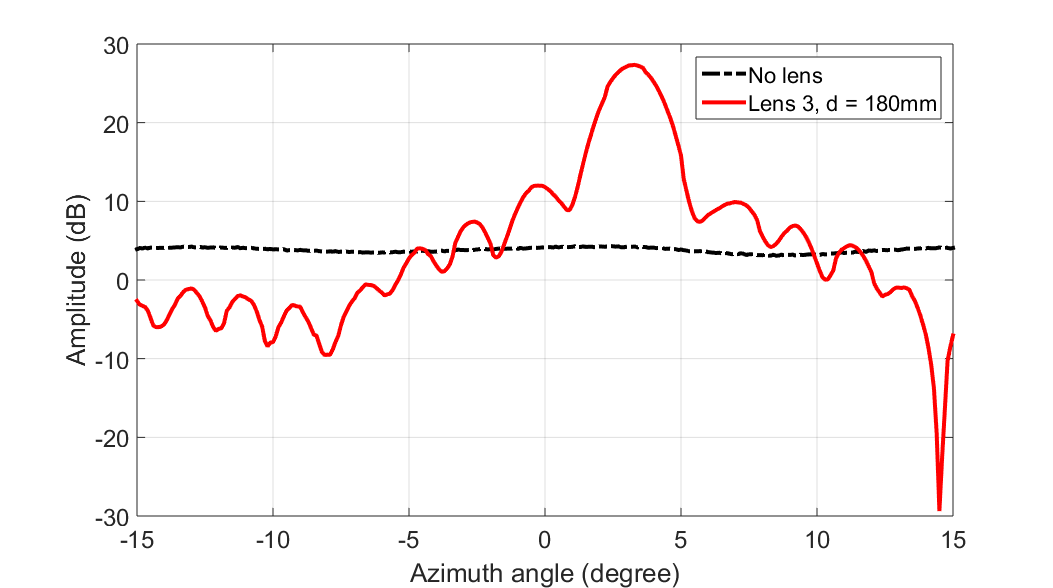}
                \caption{Azimuth radiation pattern with Lens 3}
                \label{azi3}
        \end{subfigure}
        \begin{subfigure}[b]{1.0\columnwidth}
                \centering
                \includegraphics[width=1.0\columnwidth,height = 4.5cm]{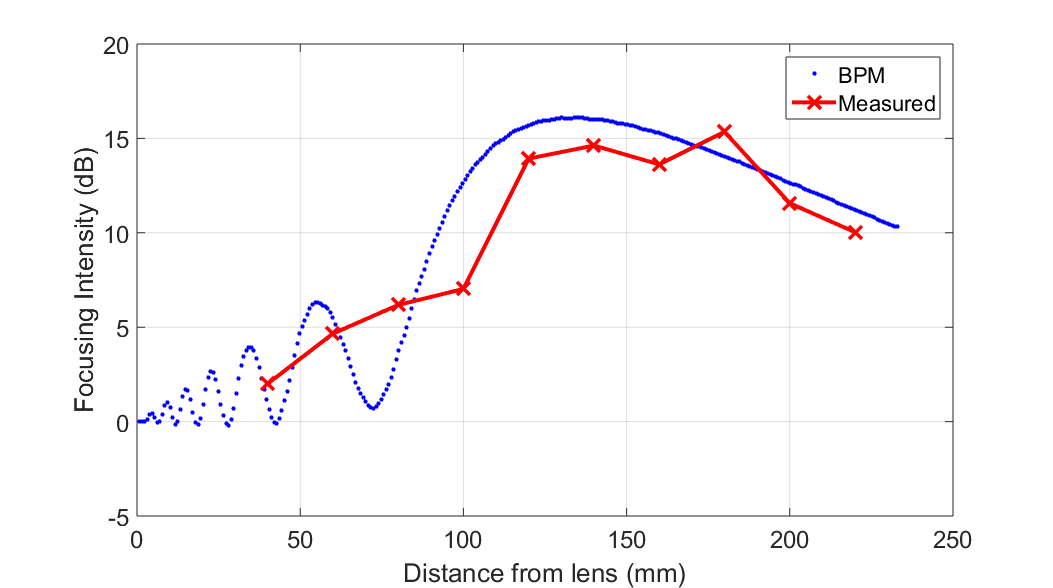}
                \caption{Axial power distribution (focusing intensity) of Lens 3}
                \label{ax3}
        \end{subfigure}
        \setlength{\belowcaptionskip}{-12pt}
        \caption{Azimuth radiation pattern and axial power distribution of three different lens antenna setups.}
        \label{Fig:lens}
\end{figure*}

\begin{figure*}[t]
       \centering
        \begin{subfigure}[b]{1.0\columnwidth}
                \centering
                \includegraphics[width=1.0\columnwidth,height = 3cm]{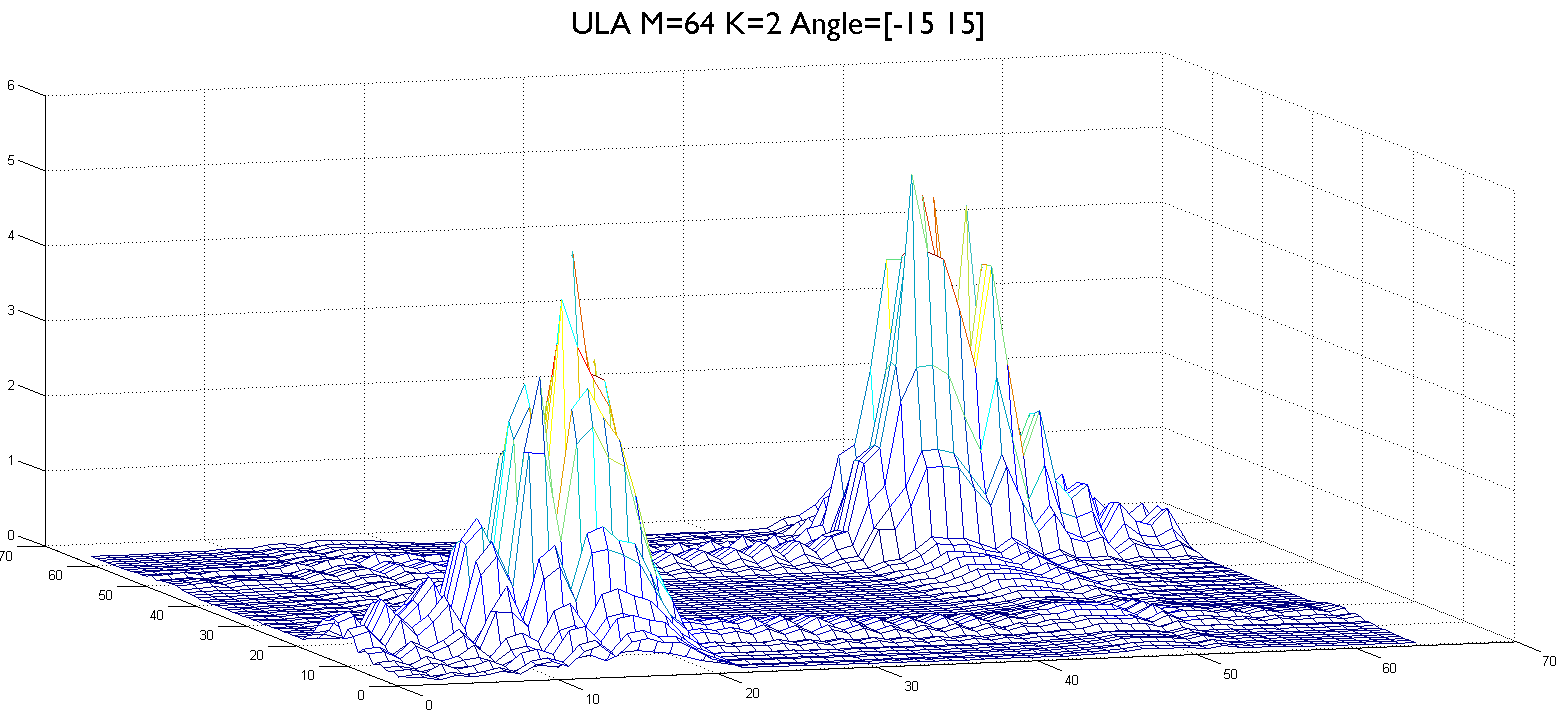}
                \caption{$\text{ULA}$, $M=64$, $K=2$, $\text{Angle}=[-15\ 15]$}
                \label{corr1}
        \end{subfigure}
        \begin{subfigure}[b]{1.0\columnwidth}
                \centering
                \includegraphics[width=1.0\columnwidth,height = 3cm]{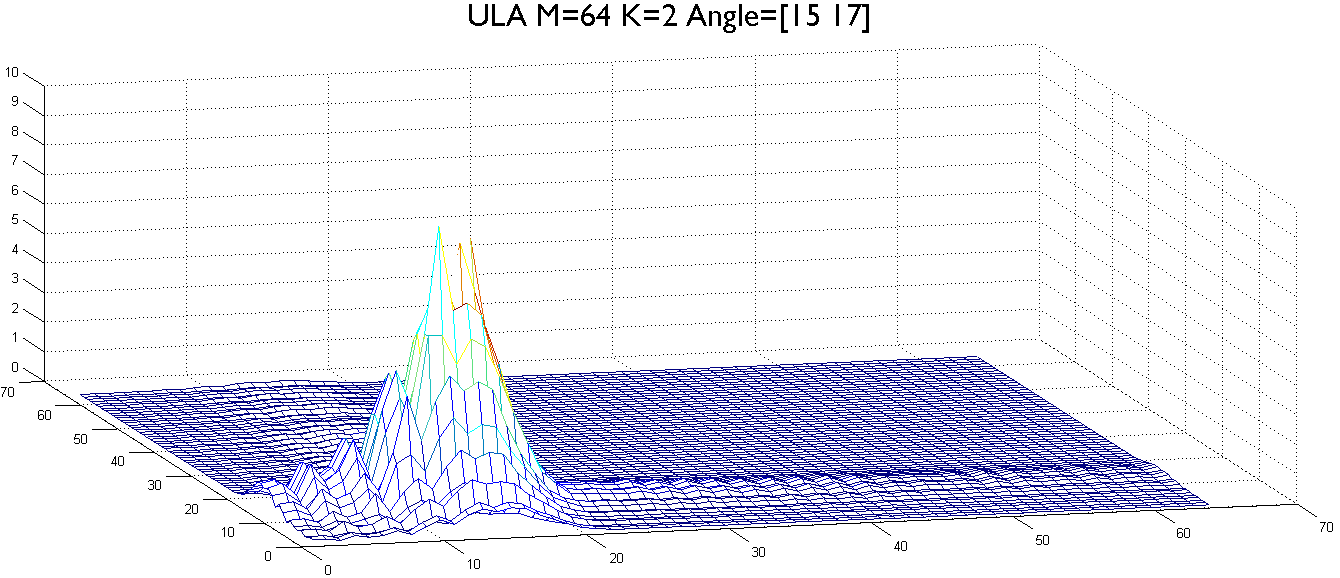}
                \caption{$\text{ULA}$, $M=64$, $K=2$, $\text{Angle}=[15\ 17]$}
                \label{corr2}
        \end{subfigure}
        \begin{subfigure}[b]{1.0\columnwidth}
                \centering
                \includegraphics[width=1.0\columnwidth,height = 3cm]{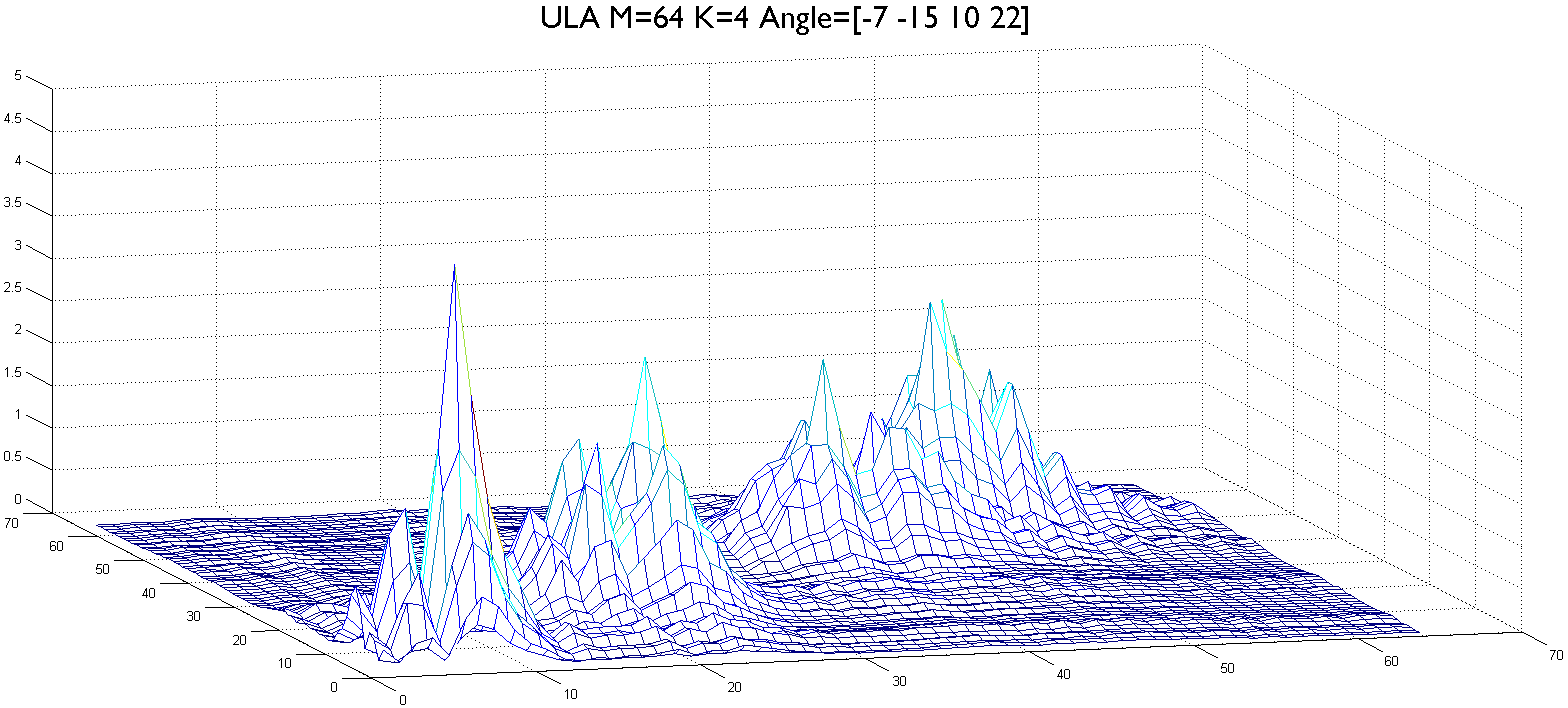}
                \caption{$\text{ULA}$, $M=64$, $K=4$, $\text{Angle}=[-15\ -7\ 10\ 22]$}
                \label{corr3}
        \end{subfigure}
        \begin{subfigure}[b]{1.0\columnwidth}
                \centering
                \includegraphics[width=1.0\columnwidth,height = 3cm]{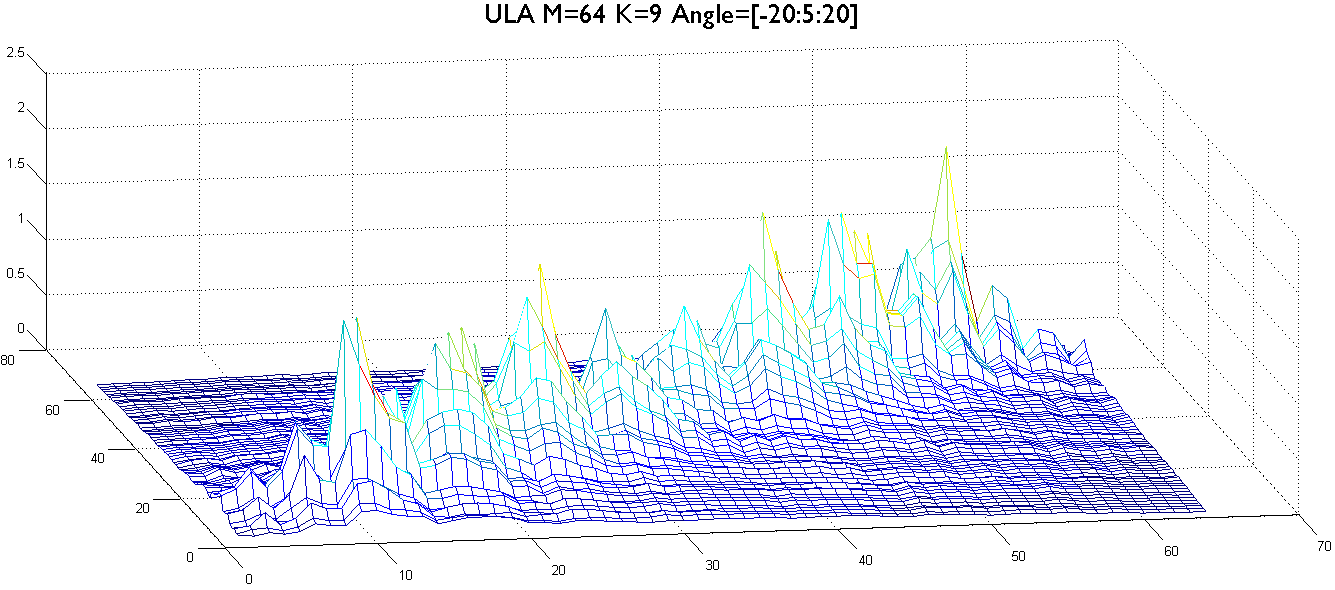}
                \caption{$\text{ULA}$, $M=64$, $K=9$, $\text{Angle}=[-20:5:20]$}
                \label{corr4}
        \end{subfigure}
        \setlength{\belowcaptionskip}{-12pt}
        \caption{Unnormalized channel correlation matrices of various user scenarios.}
        \label{Fig:corr}
\end{figure*}

\subsection{Experimental Results}
 The measurements were carried out with different lenses, different probe antenna samples, and lens separations. To measure the radiation pattern, the antenna was rotated in the range of \mbox{[-15$^\circ$,~15$^\circ$]} in both the azimuth and elevation angle domains. As the axial power distribution was measured, the separation between the lens and probe was altered, in each configuration, in 20~mm steps, resulting in an axial measurement range of 40~mm--220~mm.

 Step 1), the measurement without the RF lens, was performed to demonstrate the basic properties of the designed probe antenna samples. The $S_{11}$ parameter, in the range of 75~GHz--78~GHz in Fig.~\ref{Fig:S11}, showed a difference between the simulation and the measurements, where the center frequencies of the designed samples were slightly higher than those of the simulation. This was the expected error caused by the fabrication process since the physical dimensions of the antenna were so small that they could create a variance in the simulation result \cite{2015singlechip}. 
  
Step 2), the measurement with an RF lens, was designed to evaluate the beam characteristics in front of the RF lenses. We measured the radiation patterns for system-level performance analysis shown in Fig. 14. Figures~\ref{Fig:lens}(a), \ref{Fig:lens}(c), and \ref{Fig:lens}(e) compare the azimuth radiation patterns of the lens and no lens instances. The distance between the lens and receiving antenna was carefully determined based on the focal length of the fabricated lens. All three lenses showed an amplitude gain exceeding 20~dB, at the azimuth angle with the peak value. The lens property of focusing the power of the incident wave was thus verified, even considering the loss tangent of materials~\cite{Afsar1985}.\footnote{The beams are not exactly centered due to the misalignment of the Styrofoam jig.}  
  
The axial power distributions of the beam in front of each lens are shown in Figs.~\ref{Fig:lens}(b), \ref{Fig:lens}(d) and \ref{Fig:lens}(f). The blue one is a BPM simulation result that assumes each dielectric parabolic lens to be a thin spherical lens with an equivalent focal length. A parabolic lens was chosen over a thin spherical lens owing to a fabrication issue; since a parabolic lens has a closed form contour for both sides while a spherical lens does not. We therefore only compare the focusing intensity in these measurements. It is widely known that the thin lens approximation breaks down for the lens with a small $f/D$ ratio, but focusing intensity has comparable results with the thin lens assumption~\cite{2011elliptic}.

Possible sources of error in the measurements are numerous. These include inaccuracies of the moving Styrofoam jig, alignment problem of the transmitter, lens, and receiver, directivity of the probe antenna and standing waves between the lens, and receiver. The measurement interval at the axial measurements is 20~mm; this means that a maximum 10~mm inaccuracy could also occur. In addition, errors caused by misalignment of the horn, lens and probe antenna could be amended for each configuration. We propose these adjustments for our future work.

\section{Performance Analysis}
\label{analysis}

In this section, we evaluated the performances of a lens-embedded massive MIMO system in 2D and 3D environments.
\subsection{Performance Analysis in 2D Environments}
 We considered a single-cell downlink transmission and limited feedback scenario in which the ULA with 64 elements is separated by $d=0.5\lambda$ at the BS. The geometry of the RF lens was given as $\ell$ = $25\lambda$ and $f$ = $40\lambda$. The electric permittivity of the lens was set to 2.4. The power profile vectors vary with the angles at which the users are located, so various angles were tested. The coverage angle of the antenna array was specified as a range of [-30$^\circ$,~30$^\circ$], for the case of six equally covered sectors in a cell. We also used two precoders, ZF and MRT, and adapted various codebook designs for comparisons in both the no-lens and lens cases. Details are given for each simulation.

Figure~\ref{Fig:corr} illustrates several channel correlation matrices, $\mathbb{E}[{\tilde{\pmb{H}}}^{\ast}{{\tilde{\pmb{H}}}}]$, for various numbers of users and their angles. In Fig.~\ref{Fig:corr}(a), two energy-focused peaks were resolvable, which implies that there was a sufficient angle difference between the two users. In Fig.~\ref{Fig:corr}(b), however, the channels of two adjacent users were hard to separate at the BS, which means we need more antennas at the BS to distinguish one user from another. Figures~\ref{Fig:corr}(c) and~\ref{Fig:corr}(d) deal with multi-user cases (four and seven users, respectively). If a sufficient angle distance is guaranteed between the users, the performance of the RF lens-embedded system is not likely to be degraded significantly. From the channel correlation matrix of specific user cases, we can estimate the performance of our proposed system.

\begin{figure}[!t]
     \includegraphics[width=1.0\columnwidth]{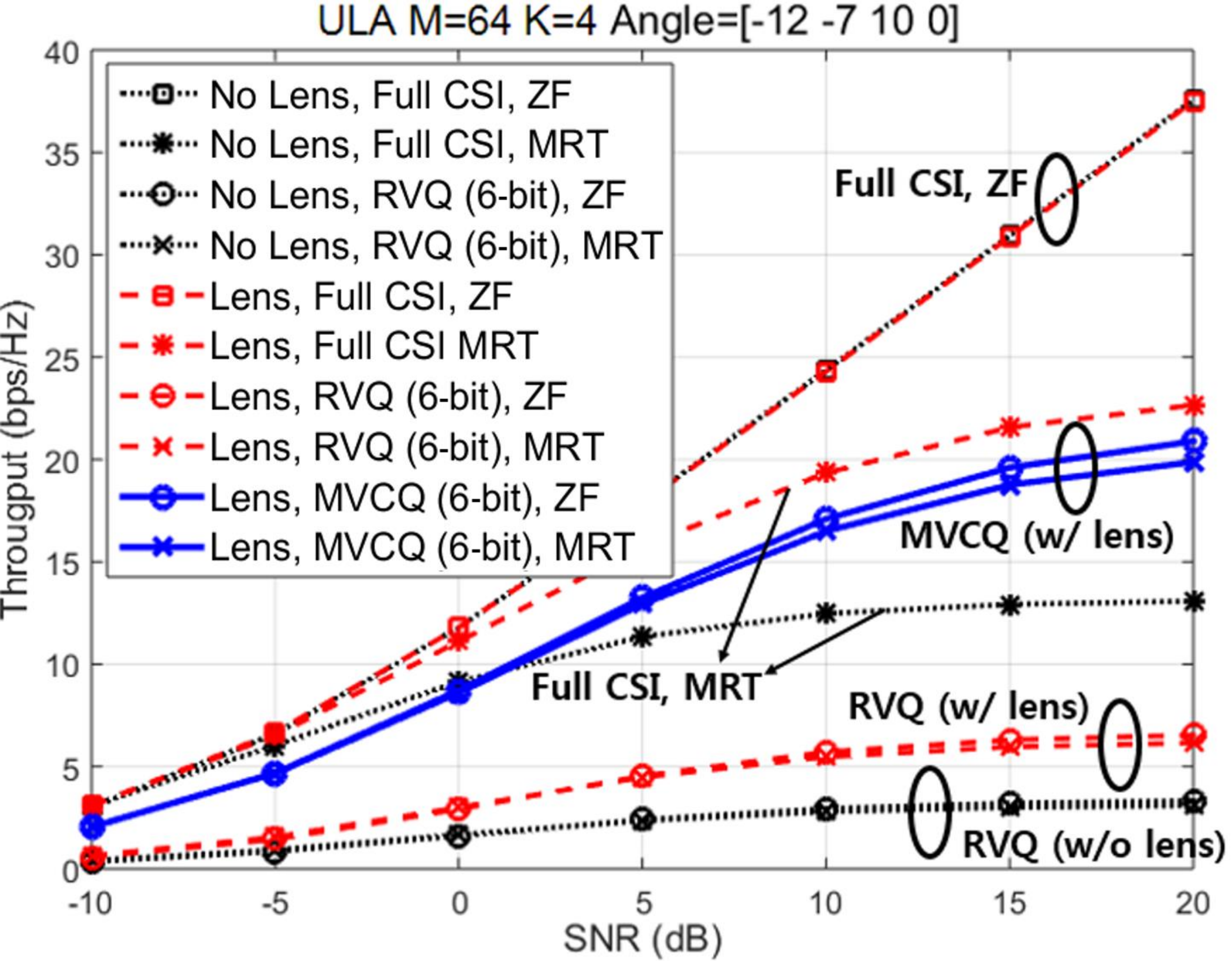}
   \caption{Achievable sum rate of massive MIMO systems with and without the RF lens.}
   \label{th1}
\end{figure}

\begin{figure}[!t]
   \centerline{\resizebox{1.0\columnwidth}{!}{\includegraphics{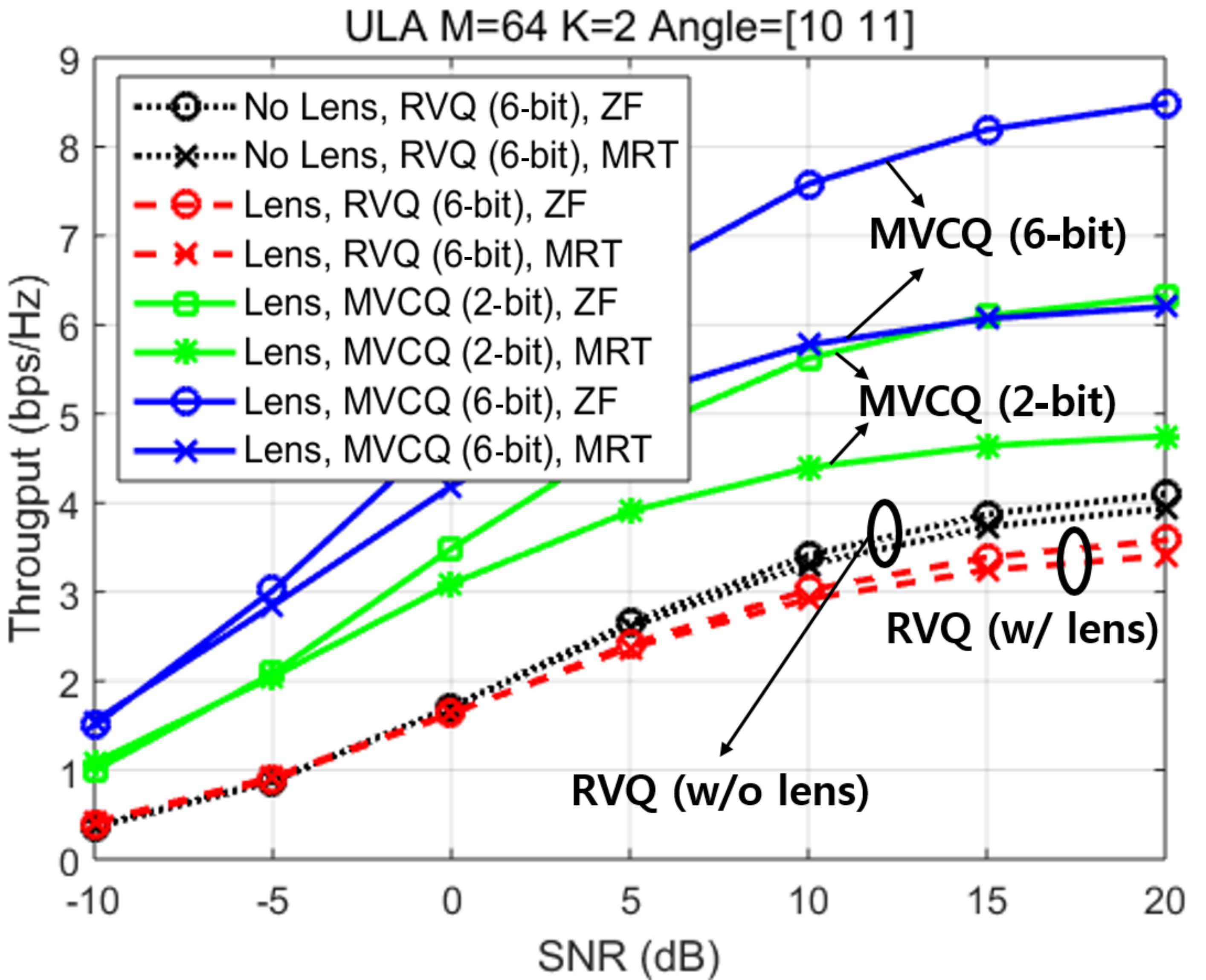}}}
   \caption{Achievable sum rate of massive MIMO systems with and without the RF lens (2-bit feedback included).}
   \label{th2}
\end{figure}

Figure~\ref{th1} plots the achievable sum rate of the massive MIMO system with and without the RF lens. Four users with -12$^\circ$, -7$^\circ$, 10$^\circ$, and 0$^\circ$ were in a cell. The BS performs with knowledge of the full CSI, or a quantized CSI used a limited feedback method employing RVQ (conventional) and MVCQ (proposed). For MVCQ codebook generation, $1\lambda$-BPM was used. If the full CSI was known to the BS, the MRT precoder would have quite a large throughput gain in the RF lens-embedded system, while the ZF precoder would remain unchanged. The performance gap of the MRT precoder between the RF lens-embedded system and the conventional system was simply understood, since the sum value of the squared channel gain increased by the energy-focusing operation of the RF lens. Regardless of the deployment of the RF lens, in limited feedback cases where the 6-bit RVQ codebook was used, no performance gap was found between the two systems in either the ZF or the MRT precoder cases. This means that the RVQ was unsuitable for RF lens-embedded massive MIMO systems. In fact, it might be even worse than no RF lens cases (see Fig.~\ref{th2}). If we consider 3D environments, we could clearly observe the gain from the lens directivity (see Fig.~\ref{slscdf}). 

The proposed codebook construction method, however, provided a significant performance enhancement in the system with the RF lens. When the MVCQ method with a 6-bit codebook was used for the channel estimation, the ZF and MRT precoders each showed a steep increase in performance. They performed comparably with the MRT precoder with the full CSI in the system with the RF lens. Both MVCQ scenarios outperformed the MRT precoder with the full CSI in the system without the RF lens in the high SNR regime. This showed that even with a quantized CSI, the proposed codebook method could achieve a similar performance to the full CSI case. In comparing the blue line (lens-embedded system, MVCQ) with the black dotted line with o and x marks (no lens, RVQ), there was excessive SNR gain between the lens-embedded system and the conventional system. The gain that goes beyond 10~dB was demonstrated for the throughput range, 2--6~bps/Hz. 

\begin{figure}[!t]
   \centerline{\resizebox{1.0\columnwidth}{!}{\includegraphics{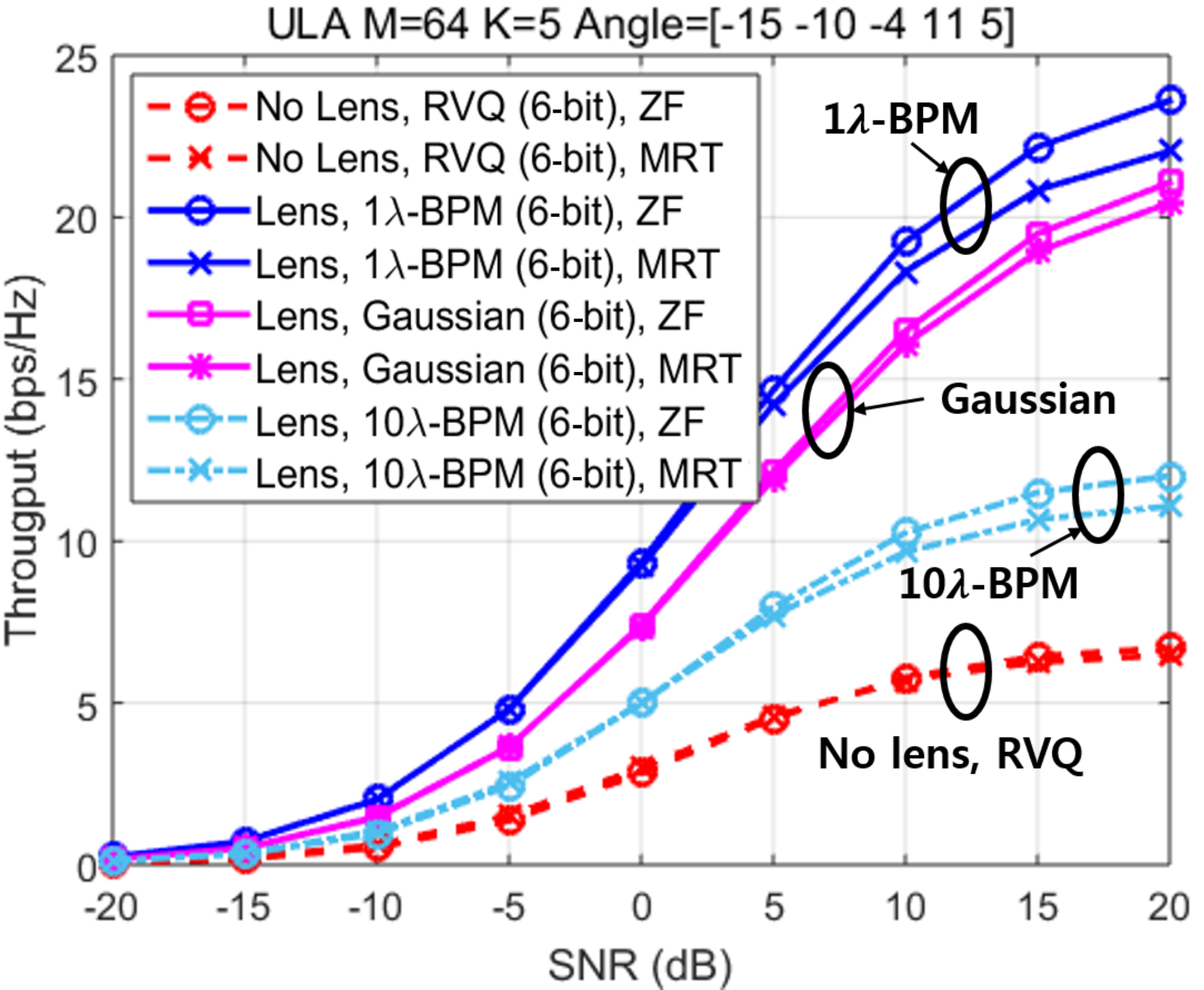}}}
   \caption{Achievable sum rate of massive MIMO systems with and without the RF lens (Low-complexity estimation for power distribution included).}
   \label{th3}
\end{figure}

\begin{figure*}[!t]
   \centerline{\resizebox{2.0\columnwidth}{!}{\includegraphics{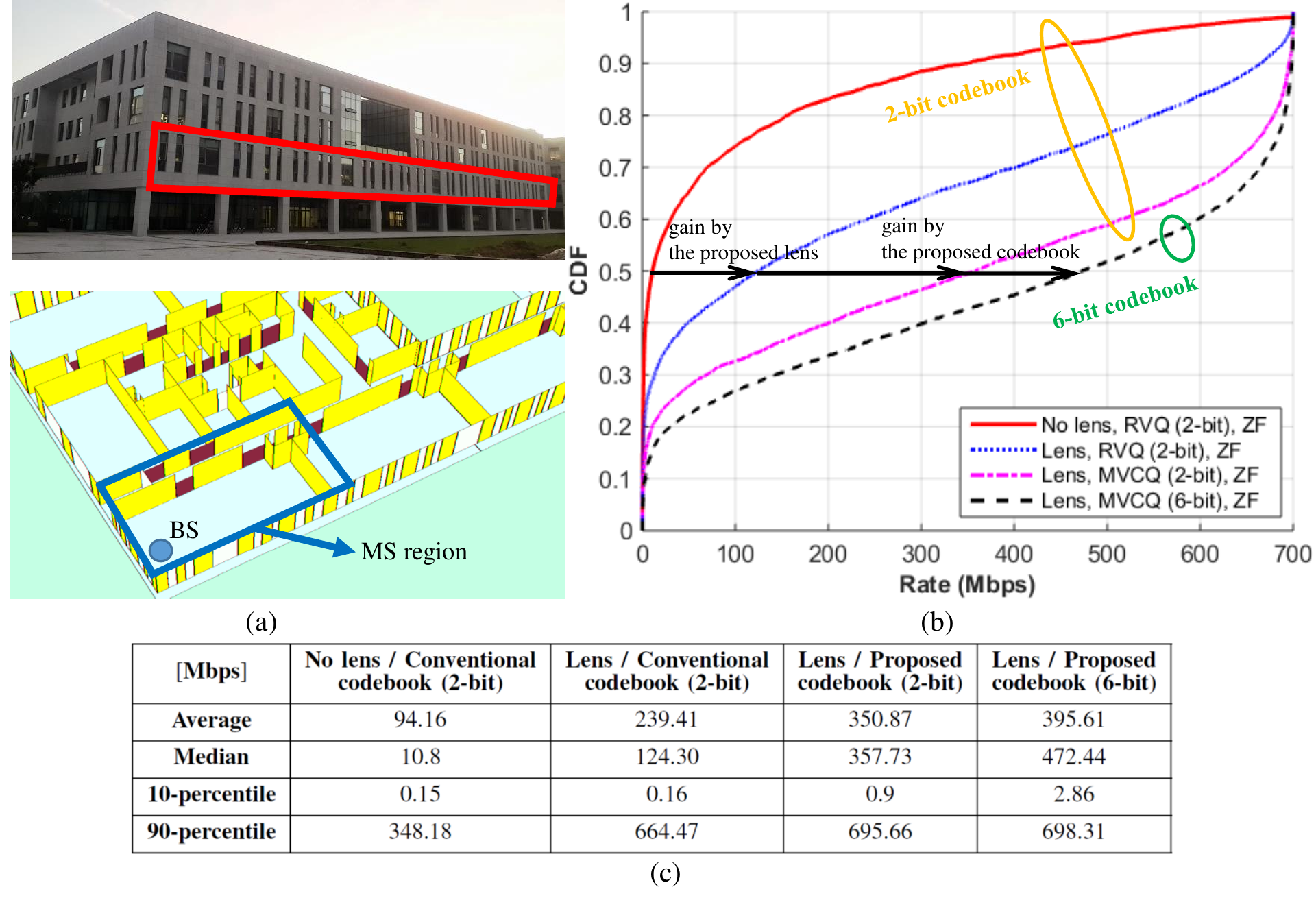}}}
   \caption{3D environments simulation result: (a) Topology for system-level performance evaluations, (b) CDF of the system throughput, and (c) median, average, and percentiles of throughput results.}
   \label{slscdf}
\end{figure*}

In Fig.~\ref{th2}, further analysis was conducted to determine the SNR gain and lower feedback overhead of the proposed codebook construction method. We considered the same system configuration as that shown in Fig.~\ref{th1} while two neighboring users were located in a single cell with angles of $10^{\circ}$ and $11^{\circ}$. For MVCQ codebook generation, $1\lambda$-BPM was used. Since the angle difference between the two users was comparatively small, the performance of the lens-embedded massive MIMO system might be degraded due to the difficulty of resolving the user signals at the BS. Additionally, the performance of the ZF precoder was far better than that of the MRT precoder, since the squared channel gain shrank due to the overlapped energy-focusing pattern. The main result showed that even though the number of feedback bits of the proposed channel feedback algorithm was 2--while that of the RVQ was 6--the ergodic achievable sum rate of the lens-embedded MVCQ system was far better than that of the RVQ system with or without the lens. For the same sum rate performance of 2~bps/Hz, 7.1~dB and 10.2~dB SNR gain were achieved for the 2-, and 6-bit MVCQ. Indeed, when the feedback bits in the MVCQ codebook increased (2-bit $\to$ 6-bit), the throughput of the proposed system using the ZF or MRT precoding method increased simultaneously to a certain level. The case where a ZF precoder was used showed more improvements in the total throughput, which was approximately 1.96~bps/Hz, while the MRT precoder case showed a slight performance gain (1.38~bps/Hz at 10~dB SNR). 

Figure~\ref{th3} compares the achievable sum rate performances of several codebook generation methods in MVCQ. We considered a situation where there were five users with -15$^\circ$, -10$^\circ$, -4$^\circ$, 11$^\circ$, and 5$^\circ$ in a cell. We included Gaussian and sub-BPM for low-complexity estimations. To specify the power coefficient based on Gaussian estimation, we used the fitted parameters given in Section IV-D. The sampling distance was given as $10\lambda$ for sub-BPM. As shown in the figure, there was performance degradation in both codebook generation methods compared with $1\lambda$-BPM. $10\lambda$-BPM was far lower than $1\lambda$-BPM in the high SNR regime, but still surpassed the RVQ case. Although the Gaussian estimation came close to $1\lambda$-BPM  and significantly reduced the codebook generation complexity, in a practical situation, it might be complicated to decide on the parameters. This is an issue for future work. 

\subsection{Performance Analysis in 3D Environments}

 We evaluated the RF lens-embedded MIMO system and the proposed codebook by using a system-level simulator based on Wireless System Engineering (WiSE)--a 3D-ray tracing tool developed by Bell Labs~\cite{valenzuela1998channel}. For realistic performance evaluations, we modeled Veritas Hall C of Yonsei University in Korea, shown in Fig.~\ref{slscdf}(a)~\cite{smallcell2015}. The BS was located at the corner of the second floor of the building with two different antennas, i.e., with/without an RF lens array. On the second floor, we uniformly distributed the MS, each of which was equipped with a single isotropic antenna. The system parameters for the evaluations included up to 256~QAM (modulation) and 100 MHz bandwidth.



 Figure~\ref{slscdf}(b) illustrates the results of the ergodic throughput of the system with four configurations. We compared a no-lens MIMO system with 2-bit RVQ (the solid line), a lens-embedded MIMO system with 2-bit RVQ (the dotted line), and a lens-embedded MIMO system with the proposed codebook method, 2-, and 6-bit MVCQ (the dashed line). Since there was no great difference between the cases with ZF and MRT precoders, we only plotted the case with the ZF precoder. Provided in the table are representative values of each case, such as average, median, 10~\%, and 90~\%.
 
  The result implies that the throughputs of the lens-embedded massive MIMO systems increased with the help of two factors: the directivity of the RF lens and suitable codebook adaptation. The average throughput of the lens-embedded MIMO system was increased by approximately 254~$\%$ compared to that of a conventional MIMO system. When MVCQ method was performed in a lens-embedded MIMO system, we observed more significant gain in the throughput performances. For the lens-embedded system with MVCQ, about 373~\% (2-bit) and 420~\% (6-bit) improvements were observed in the average throughput performance. Further dramatic improvements were shown in median values. 
  
\section{Conclusion}
\label{conclusion}
In this paper, the RF lens-embedded massive MIMO system was found to significantly reduce the feedback overhead in conventional massive MIMO systems. We adopted BPM to calculate the propagation and the focusing of the incident beams controlled by the RF lens. To verify the accuracy of BPM, we fabricated an RF lens operating at mmWave and compared the theoretical results with measurement data. 
BPM was also utilized in the construction of our new channel model. A new channel quantization-based codebook method, MVCQ, was proposed to generate adaptive codebooks that could reflect multi-channel variances. Analytical and numerical results confirmed that the proposed MVCQ codebook with only a few bits achieved significant performance throughput gain in realistic environments.

\section*{Acknowledgment}
The authors would like to thank Dr. B.-N. Kim, Mr. M. Jeong, and Mr. C. Kim for their helpful discussions.

\renewcommand{\baselinestretch}{1.0}
\bibliographystyle{IEEEtran}
\renewcommand{\baselinestretch}{1.0}
\bibliography{references_MVCQ}

\end{document}